\documentclass[superscriptaddress,amsmath,amssymb,aps,prb,twocolumn,floatfix]{revtex4-2}

\usepackage{graphicx}
\usepackage{physics}
\usepackage{bm}
\usepackage[colorlinks, linkcolor=mycol, citecolor=mycol, urlcolor=mycol, breaklinks]{hyperref}
\usepackage{braket}
\usepackage{bbold}
\usepackage{bm}
\usepackage{ulem}
\usepackage{tabularx, colortbl}
\usepackage{comment}
\usepackage{appendix}

\definecolor{mycol}{RGB}{10,55,130}
\definecolor{mycol2}{RGB}{50,187,108}
\definecolor{mycol3}{RGB}{204,102,0}

\begin{document}
\title{Engineered dissipation induced entanglement transition in quantum spin chains: from logarithmic growth to area law}

\author{T.~Botzung}
\affiliation{Institute for Quantum Information, RWTH Aachen University, D-52056 Aachen, Germany}

\affiliation{
Peter Gr\"unberg Institute, Theoretical Nanoelectronics, Forschungszentrum J\"ulich, D-52425 J\"ulich, Germany}

\author{S.~Diehl}
\affiliation{Institut  f\"ur Theoretische Physik, Universit\"at zu K\"oln, D-50937 Cologne, Germany}

\author{M.~M\"uller}
\affiliation{Institute for Quantum Information, RWTH Aachen University, D-52056 Aachen, Germany}

\affiliation{
Peter Gr\"unberg Institute, Theoretical Nanoelectronics, Forschungszentrum J\"ulich, D-52425 J\"ulich, Germany}

\date{\today}

\begin{abstract}

Recent theoretical work has shown that the competition between coherent unitary dynamics and stochastic measurements, performed by the environment, along wavefunction trajectories can give rise to transitions in the entanglement scaling. In this work, complementary to these previous studies, we analyze a situation where the role of Hamiltonian and dissipative dynamics is reversed. We consider an engineered dissipation, which stabilizes an entangled phase of a quantum spin$-1/2$ chain, while competing single-particle or interacting Hamiltonian dynamics induce a disentangled phase. Focusing on the single-particle unitary dynamics, we find that the system undergoes an entanglement transition from a logarithmic growth to an area law when the competition ratio between the unitary evolution and the non-unitary dynamics increases.  We evidence that the transition manifests itself in state-dependent observables at a finite competition ratio for Hamiltonian and measurement dynamics. On the other hand, it is absent in trajectory-averaged steady-state dynamics, governed by a Lindblad master equation: although purely dissipative dynamics stabilizes an entangled state, for any non-vanishing Hamiltonian contribution  the system ends up irremediably in a disordered phase. In addition, a single trajectory analysis reveals that the distribution of the entanglement entropy constitutes an efficient indicator of the transition. Complementarily, we explore the competition of the dissipation with coherent dynamics generated by an interacting Hamiltonian, and demonstrate that the entanglement transition also occurs in this second model. Our results suggest that this type of transition takes place for a broader class of Hamiltonians, underlining its robustness in monitored open quantum many-body systems.
\end{abstract}

\maketitle


\section{ Introduction }
\label{sec0}

 Out-of-equilibrium quantum dynamics is a fundamental scientific challenge appearing in many aspects of modern physics. Several works have explored out-of-equilibrium phenomena in \textit{isolated} systems with remarkable developments, including thermalization of closed quantum systems~\cite{Rigol2008, D'Alessio2016}, many-body localization (MBL)~\cite{Abanin2019, Basko2006},  quantum chaos~\cite{Maldacena2016, D'Alessio2016}, Floquet time-crystals~\cite{Khemani2019}, or quantum simulation~\cite{Georgescu2014}. In parallel, the dynamics of \textit{open} quantum systems offers  novel perspectives for out-of-equilibrium phenomena.  The signatures of the competition between unitary and non-unitary dynamics range from radiative decay in driven open quantum systems~\cite{Mollow1975, Dicke1998}, phonon-induced decay in solid-state devices~\cite{Golovach2004, Sukachev2017}, nonequilibrium phase transitions in driven open quantum systems~\cite{Diehl2010, Eisert2010, Lee2011, Kessler2012, Honing2012, Horstmann2013, Lee2013, Lang2015Jul, Jin2016, Maghrebi2016, Sieberer2016, Foss-Feig2017, Minganti2018}, or engineering of dissipative quantum states~\cite{Poyatos1996, Diehl2008, Verstraete2009, Krauter2011, Barreiro2011, Schindler2013}. In open quantum systems, due to the coupling between the quantum system and the environment, information about the state of the system continuously leaks into the environment.  Conditioned on the access to the environment, information about the state of the system can be retrieved. In that case, it is possible to describe the system dynamics in the form of quantum trajectories (QT), which correspond to individual pure state wavefunctions that evolve conditional on the monitoring realized by the environment~\cite{Wiseman1996, Warszawski2002Dec, Warszawski2002Dec2}. The density matrix description, governed by a Lindblad master equation, is recovered by averaging over an ensemble of such trajectories, corresponding to unread measurements.

In the QT context, when both coherent unitary and measurement dynamics compete, the study of the dynamics of entanglement has led to the discovery of novel measurement-induced phase transitions (MITs), also referred to as entanglement phase transitions (EPTs). They are characterized, for example, by a qualitative change in the scaling law of the entanglement entropy. A prime instance of a MIT was recently found in open-system dynamics of quasilocal random unitary quantum circuits interleaved with local projective measurements~\cite{Li2018, Chan2019, Skinner2019, Li2019, Jian2020, Bao2020}. Here, the unitary dynamics tends to generate a volume law entanglement, i.e., a linear growth of entanglement with the system size. In stark contrast, measurements typically destroy entanglement, as they locally collapse the system wavefunction, leading to an area law phase, where the entanglement saturates. Following these advances, various theoretical~\cite{Jian2020, Bao2020} and numerical investigations have aimed at understanding the diverse facets of the volume-to-area law EPT in similar hybrid quantum dynamics~\cite{Szyniszewski2019, Zabalo2020, Gullans2020Aug, Gullans2020Oct, Choi2020}. In more generalized scenarios, some studies have provided evidence for enticing novel MITs, such as, e.g., in random tree tensor networks~\cite{Nahum2021}, in symmetric random quantum circuits~\cite{Sang2020Apr, Bao2021Feb, Lavasani2021Mar}, and in spacetime dual nonunitary circuits~\cite{Ippoliti2021Mar}.

Triggered by this discovery, several subsequent theoretical studies have identified a similar transition in other physical models. A volume-to-area law transition  was found to occur in a one-dimensional Bose-Hubbard model subject to random projective measurements~\cite{Tang2020}, or two-body losses~\cite{Goto2020}, and in a spin chain subject to continuous position measurements~\cite{Fuji2020Aug, Fuji2021Feb}. In addition, the analysis of a chain of free fermions under local continuous monitoring~\cite{Cao2019} has revealed a different MIT from a critical phase with a logarithmic scaling  of the entanglement entropy to an area law~\cite{Alberton2021}. Very recently, MITs were found in fermionic models with either long-range non-unitary evolution~\cite{Minato2021Apr} or long-range Hamiltonian dynamics~\cite{, Zhang2021May, Muller2021May}. 

In the aforementioned studies, only hermitian measurement operators, and thus the destructive character of the monitoring, were considered. However,  engineering dissipation via non-hermitian generalized measurement operators can also generate quantum correlations. Indeed, it has been shown both  theoretically~\cite{Diehl2008, Kraus2008, Verstraete2009, Muller2012Jul} and experimentally~\cite{Krauter2011, Barreiro2011, Schindler2013, Lin2013Dec} that the dissipation can be appropriately designed to prepare and stabilize correlated quantum many-body states. 
For instance, it has been demonstrated that long-range phase coherence can be realized by an iteration of purely dissipative quantum processes in bosonic systems, creating entangled phases~\cite{Diehl2008, Kraus2008, Schindler2013}. Moreover, a wide variety of many-body states can be prepared by dissipation engineering. Examples include, among others, topological phases in spin systems~\cite{Weimer2010May}, and in fermionic ones~\cite{Diehl2011Dec, Bardyn2012Sep, Budich2015Apr}, pairing states for fermions in optical lattices~\cite{Yi2012May}, and the emergence of a dissipation-stabilized phase in a  quantum gas~\cite{Ferri2021Apr}.
In particular, the generation of entangled states via dissipation has been investigated in a number of quantum optical and solid state systems, ranging from cavity QED~\cite{Plenio1999, Clark2003, Vacanti2009, Kastoryano2011, Busch2011, Tomadin2012Jun, Reiter2012, Li2017}, ion traps~\cite{Poyatos1996, Cho2011, Barreiro2011, Mueller2011, Bermudez2013Mar, Reiter2016}, optical lattices~\cite{Verstraete2009, Kordas2012Nov}, Majorana qubits~\cite{Gau2020, Gau2020Oct}, atomic ensembles~\cite{Krauter2011}, superconducting qubits~\cite{Leghtas2013Aug, Shankar2013Dec}, to name a few.

This ability to tailor in particular non-trivial dissipative processes gives rise to new scenarios of competition between unitary and dissipative dynamics. In the context of MITs, very recently, EPTs were found for example in measurement-only models where the competition arises from two incompatible dissipative channels~\cite{VanRegemortel2020} or from different projective measurements~\cite{Ippoliti2021}. In contrast, in this work we study a monitored quantum many-body system where the dissipation generates an entangled state, and competes with a coherent dynamics detrimental to the creation of entanglement. In light of that perspective, it begs the question whether we can find a similar entanglement transition in the measurement dynamics of a broader class of open quantum systems than the ones studied so far.

\subsection{Key results}
\paragraph{Models.---}This work addresses these questions by studying the quantum trajectory dynamics of a chain of spin-$1/2$ particles subject to coherent dynamics and engineered dissipation. The engineered dissipation, for its part, symmetrically delocalizes particles over pairs of neighboring sites.  For the coherent dynamics, we consider two cases, a single-particle staggered potential and a second interacting model with nearest-neighbour interaction between spin excitations. 

\paragraph{Numerical approach.---} The simulation of the dynamics of our models  represents a numerical challenge for large system sizes. To cope with this, we combine the wave-function-based Monte Carlo approach~\cite{Dalibard1992, Carmichael1993, Molmer1993} with an efficient matrix product state (MPS) and time-evolving block decimation (TEBD) algorithm~\cite{Vidal2004, Verstraete2008, Schollwock2011, Paeckel2019}. In our analysis, this method enables us to simulate the dynamics of a chain of up to 80 sites, which is required to distinguish between scaling laws, e.g. of the entanglement entropy. 
\paragraph{Entanglement phase transition.---} Using this numerical tool, we carry out an extensive analysis of the entanglement entropy (EE). We demonstrate, for both models, that by increasing the competition between unitary and dissipative dynamics, the EE changes from a logarithmic scaling law to an area law at a sharply defined point, indicating an EPT. This fact shows that even a single-body Hamiltonian is able to induce this transition.
\paragraph{State-dependent observables and higher order correlation functions.---} Particular attention has to be paid to the way the ensemble average is performed. Indeed, these EPTs manifest themselves only in the dynamics of individual quantum (measurement) trajectories. In other words, only the ensemble average of non-linear functions of the state vector of individual trajectories, such as the entanglement entropy, do exhibit the transition. In stark contrast, linear functions of the state occult the transition. We attest this statement by showing that the transition is absent in trajectory-averaged steady-state dynamics, governed by a Lindblad equation: there, an infinitesimal Hamiltonian contribution to the  system is sufficient to drive the system irremediably into a disordered phase. On the contrary, we show that higher-order correlation functions (non-linear functions)  witness the transition at a finite competition ratio between coherent Hamiltonian and dissipative dynamics, and thus constitute an excellent alternative probe of the EPT. 

\paragraph{Single trajectory analysis.---} 
Complementary to the entanglement signatures that are visible in the trajectory ensemble, we establish that the temporal distribution of the EE also exhibits the transition. By introducing a new detailed analysis at the level of a single trajectory, we also evidence a self-averaging property of the EE dynamics. More precisely, at a small competition ratio, the EE stays close to the EE of the purely dissipative steady-state and therefore follows a unimodal distribution. In contrast, for a large competition ratio, the system along a trajectory spends time between zero (weakly) entangled states and a state with a finite value of EE, which leads to a bimodal distribution. We find that the behavior of the standard deviation of this distribution reveals the change in its behavior, and ultimately the EPT. Importantly, this behavior is already present in small chains of $20-30$ spins. This shows that the analysis of the standard deviation of the EE outperforms the usual method based on a resource-intensive finite-size scaling analysis of the EE. This uni-to-bi modal behavior is present in the two cases of coherent dynamics we have considered. The presence of this transition with similar characteristics in both models suggests that the EPT is robust and present for a wide range of driven-dissipative open systems.

\subsection{Structure of the paper}

The remainder of the paper is organized as follows. In Sec.~\ref{sec1}, we introduce our model from the perspective of a quantum-jump approach. In Sec.~\ref{sec2}, we show the numerical results for the EE illustrating an EPT and then map out the phase diagram of our first model. In Sec.~\ref{sec3}, a detailed analysis of the single-particle correlation functions and their square is realized. In Sec.~\ref{sec4}, we illustrate the action of the jump operators on the correlations and the EE during a single trajectory.  In addition, we provide a  statistical analysis of the temporal distribution of the entropy along a single long trajectory. In Sec.~\ref{sec5} we analyse the situation with an alternative competing Hamiltonian of interacting spins. By studying the EE, we evidence an EPT, and we establish the critical point of the transition. We conclude in Sec.~\ref{sec: Discussion}.
Further details on the numerical methodology and analysis are provided in the Appendices.

\section{Model}
\label{sec1}

\subsection{Quantum jump trajectory approach}

Our study is based on the quantum trajectory approach, which describes the evolution of a state vector $\ket{\psi^{m}(t)}$ conditioned by a particular sequence of jump (or detection) events or their absence. In this framework, each realization $m$ provides a different quantum trajectory in Hilbert space, where the action of the environment (or measurement) on the system is realized via random jump processes. Information about the system's state are then obtained by stroboscopic measurement of the environment. Importantly, these measurements have to be realized within a time scale that is on the one hand sufficiently short to capture the system dynamics and on the other hand larger than the correlation time of the environment. In quantum optics, the trajectory dynamics has a simple physical interpretation, where for instance jump processes act as spontaneous emission into the environment. In this case, each event scatters photons that could, in principle, all be measured. Knowing if a jump has occurred or not gives us information about the state of the system and is relevant in studies of photon counting~\cite{Carmichael1989, Tian1992, Wiseman1993, Plenio1998} or continuous measurements~\cite{Dum1992, Barchielli1993, Srinivas1996}. 

Under these conditions, the evolution of a quantum trajectory for an infinitesimal time step $\textrm{d}t$ can correspond to two situations:

(i) during the interval $[t, t +  \textrm{d}t]$ no jump (or measurement) occurs with a probability given by
\begin{equation}
    \label{eq: proba_no_jump}
    \begin{array}{cl}
    P_{\textrm{no jump}} & \equiv 1 -  \textrm{d}t \sum_{\mu} \bra{\psi^{m}(t)} \kappa_{\mu} L_{\mu}^{\dagger} L_{\mu} \ket{\psi^{m}(t)}, \\
    & \equiv 1 -  \textrm{d}t \sum_{\mu} \delta P_{\mu}
    \end{array}
\end{equation}
where $L_{\mu}$ are  jump operators that describe general dissipative dynamics, including, e.g., measurement decoherence and loss processes. $\delta P_{\mu}$ corresponds to the probability that the specific operator $L_{\mu}$ will act during this time step $\textrm{d}t$.  Then, the system evolves through an effective Hamiltonian $H_{\textrm{eff}} = H_0 - i/2 \sum_{\mu }\kappa_{\mu} L_{\mu}^{\dagger} L_{\mu}$, i.e. 
\begin{equation}
\label{eq: evolution_effective}
    \ket{\psi^{m}( t + \textrm{d}t)} = e^{-iH_{\textrm{eff}}  \textrm{d}t}\ket{\psi^{m}(t)}, 
\end{equation}
where the Hamiltonian $H_0$  generates coherent unitary dynamics and $\kappa_{\mu}$ is the dissipative rate. There are few things to note here. First, the probability Eq.~\eqref{eq: proba_no_jump} represents a gain of information about the state of the system, since we know that no detection event has been recorded. The latter coincides notably with a continuous measurement description. Second, the non-hermiticity of $H_{\textrm{eff}}$  causes a decay of the norm of the state $\ket{\psi^{m}(t)}$, which imposes a re-normalization of the state after step~\eqref{eq: evolution_effective}.

(ii) The second situation represents the recording of a jump at a certain time $t$. These events appear randomly with probability
\begin{equation}
    \label{eq: proba_jump}
     P_{\textrm{jump}} \equiv  \textrm{d}t \sum_{\mu} \bra{\psi^{m}(t)} \kappa_{\mu} L_{\mu}^{\dagger} L_{\mu} \ket{\psi^{m}(t)}.
\end{equation}
The state vector of the system after a particular event $\mu$ reads
\begin{equation}
    \ket{\psi^{m}(t +  \textrm{d}t)} = \frac{ L_{\mu} \ket{\psi^{m}(t)} }
    { \sqrt{ \bra{\psi^{m}(t)} L_{\mu}^{\dagger} L_{\mu} \ket{\psi^{m}(t)} } }.
\end{equation}
The particular event $\mu$ is chosen from all the possible $\mu$ via the probability $\delta P_{\mu}/P_{\textrm{jump}}.$ 

 In general, due to the stochastic nature of the measurement dynamics of a pure state trajectory, we need to average over an ensemble of many trajectories to extract information independent of this randomness. Note that this type of average naturally comes out if we do not have access to the environment degree of freedom. Indeed, taking the average over many trajectories leads back to the Lindblad master equation: 
\begin{equation}
    \label{eq:mastereq}
    \dot{\rho} = -i [H_0, \rho] +  \sum_l \kappa_l \big( 2 L_{\mu} \rho L_{\mu}^{\dagger} - L_{\mu}^{\dagger} L_{\mu} \rho -\rho L_{\mu}^{\dagger}  L_{\mu} \big), 
\end{equation}
which describes general open quantum system dynamics. Nevertheless, we must pay attention when we take the ensemble average. In particular, if we consider the general functional form $\mathcal{F}[\rho^{m}_t]$ of $\rho^{m}_t = \ket{\psi^{m}(t)} \bra{\psi^{m}(t)}$, we have the important distinction
\begin{equation}
\begin{array}{ccl}
    \overline{\mathcal{F}[\rho^{m}_t]} & = & \mathcal{F}[\overline{\rho^{m}_t}], \quad  \text{if $\mathcal{F}$ is linear in $\rho^{m}_t$}, \\
    & & \\
     \overline{\mathcal{F}[\rho^{m}_t]} & \neq & \mathcal{F}[\overline{\rho^{m}_t}], \quad \text{if $\mathcal{F}$ is \textit{nonlinear} in $\rho^{m}_t$},
\end{array}
\end{equation}
where here $\overline{\mathcal{F}}$ denotes the average over the trajectories ensemble, e.g. the mean state reads $\rho_t = \overline{\rho^{m}_t} = \frac{1}{\mathcal{M}} \sum_{m=1}^{\mathcal{M}} \rho^{m}_t$, where $\mathcal{M}$ is the number of trajectories. The averages of linear functions of the state correspond, in the limit of $\mathcal{M} \to \infty$, to a density matrix description governed by the Lindblad master equation which masks the transition in the trajectory ensemble. However, the averages of non-linear functions of the state will witness a transition in the trajectory ensemble~\cite{Alberton2021, Buchhold2021}. In what follows, we denote the average expectation value of any operator $\hat{O}$ at a time $t$ by $\overline{\hat{O}_t}$,
\begin{equation}
 \overline{\hat{O}_t} = \frac{1}{\mathcal{M}} \sum_{m=1}^{\mathcal{M}} \bra{\psi^{m}(t)} \hat{O} \ket{\psi^{m}(t)}.
\end{equation}

\subsection{Staggered potential and engineered dissipation}

\begin{figure}[t!]
    \centering
    \includegraphics[width=1\columnwidth]{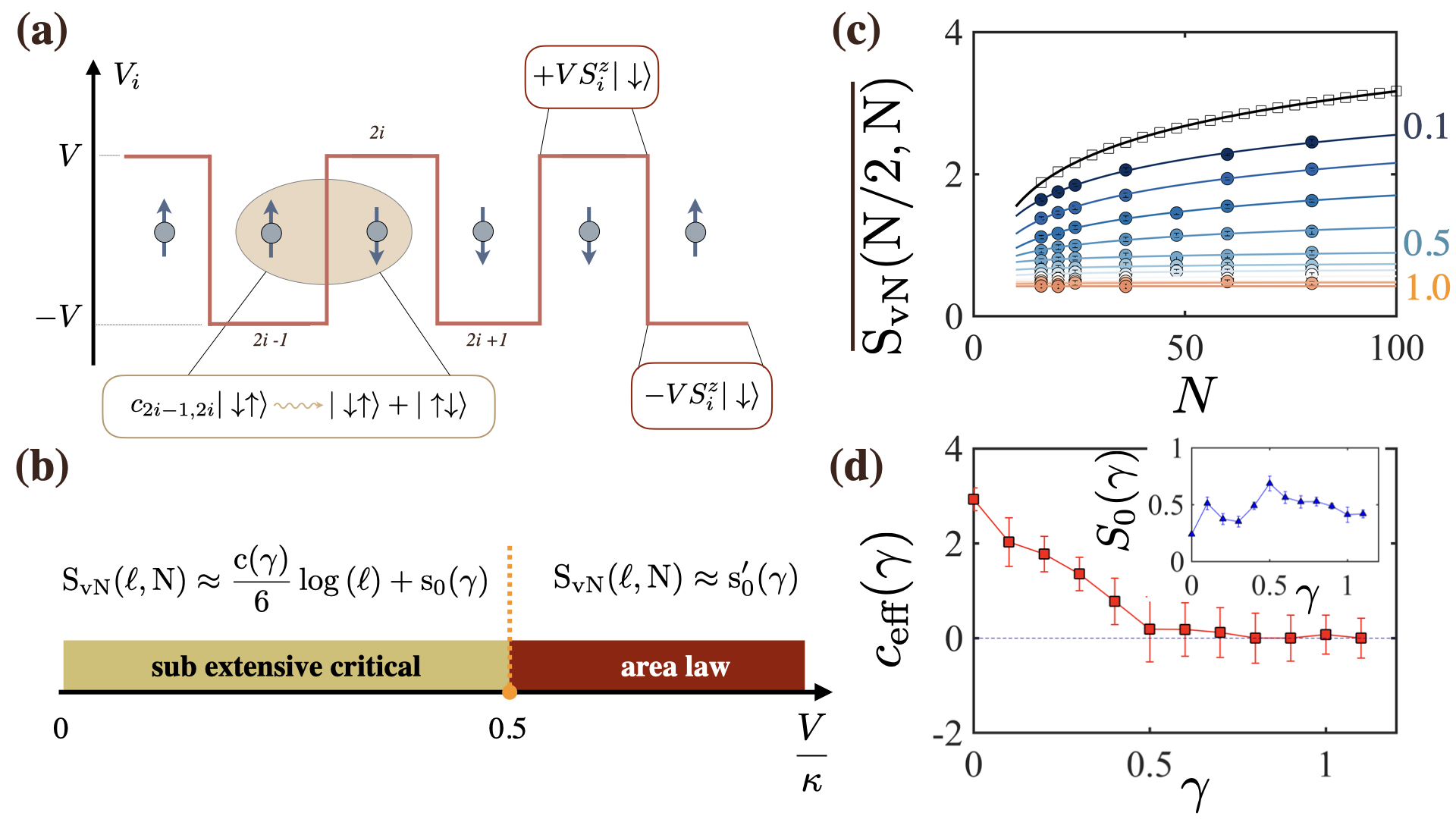}
    \caption{(a) A chain of spins subjected to a staggered potential and jump operators of the form Eq.~\eqref{eq:jumpope}. (b) Schematic phase diagram showing the different regimes of entanglement scaling as a function of the ratio $\gamma = V/\kappa$. (c) Scaling of the half-chain von Neumann entropy for various dimensionless ratio $\gamma$ (see the color code). The black line is the exact entropy of the Dicke state $\ket{D_N^{(k=0.25N)}}$, obtained from Eq.~\eqref{eq: EE_dicke}.  We observe a logarithmic scaling at a low ratio, while a high competition ratio leads to an area law scaling. (d) The effective central charge and residual entropy (inset) extracted by fitting the data in (c) with Eq.~\eqref{eq: h_CC}. It shows a transition around $\gamma_c \approx 0.5$.}
    \label{fig: fig1}
\end{figure}

We  study a chain of $N$ spins-$1/2$, subject to a staggered potential, described by the Hamiltonian
\begin{equation}
    H_0 = V \sum_i^{N} (-1)^{i} \sigma^{z}_i, 
\end{equation}
where $\sigma^{z}$ denotes the local Pauli matrix and $V$ the strength of the potential. For the dissipation, we consider the jump operators $c_l$,
\begin{equation}
\label{eq:jumpope}
    c_l = c_{\langle i, j \rangle} = (\sigma_i^{+} + \sigma_{j}^{+} ) ( \sigma_i^{-} - \sigma_{j}^{-} ), 
\end{equation}
where $\sigma_i^{\pm} = (\sigma_i^{x} \pm i\sigma_i^{y} )/2$ are spin-1/2 raising and lowering operators acting on spin $i$ and with a spatially homogenous dissipative rate $\kappa_l \equiv \kappa \equiv 1 $. Each of these operators acts on a pair of adjacent lattice sites $l \equiv \langle i, j \rangle$. A single spin excitation is symmetrically delocalized over the two sites, as visualized in Fig.~\ref{fig: fig1} (a). Note that the dissipative dynamics generated by the set of operators (\ref{eq:jumpope}) conserves the total number of excitations or total magnetization $\sum_{i=1}^N \sigma_i^z$ of the system.

The ensemble averaged dynamics corresponds to a driven dissipative model, 
\begin{equation}
    \label{eq:mastereq_eff}
    \dot{\rho} = -i [H_{\text{eff}}, \rho] +  \kappa \sum_l \big( c_l \rho c_l^{\dagger}), 
\end{equation}
where 
\begin{equation}
\label{eq: Heff}
H_{\text{eff}} = H_0 - \frac{i\kappa}{2}\sum_l  c_l^{\dagger} c_l.
\end{equation}
In several works~\cite{Diehl2008, Schindler2013}, it has been shown that these types of dissipative processes can be used to prepare a quantum system in a pure state with long-range phase coherence. Indeed, the symmetric phase-locking on each pair of sites generates phase coherence over the whole system, and attracts the system to a unique   dynamical fixed point or,  pure dark state. More specifically, within the subspace of $k$ excitations on a chain of $N$ spins, this pure dark state is given by the Dicke state

\begin{equation}
\label{eq: Dicke_state}
\ket{D_N^{(k)}} = (k!) \binom{N}{k}^{-\frac{1}{2}} \Big(\sum_{i=1}^{N} \sigma_{i}^{+} \Big)^{k} \ket{\downarrow}^{\otimes N}.
\end{equation}
The Dicke state is characterized by quantum mechanical off-diagonal long-range order. In fact, in the absence of coherent dynamics, we expect that this state is within the $k$-excitation subspace the unique steady state of the master equation~\eqref{eq:mastereq_eff} and any initially mixed state will evolve into $\ket{D_N^{(k)}}$ (numerical details can be found in Appendix~\ref{App: ConvDickeState}).

In the following, we investigate the competition between the coherent and dissipative dynamics, which is quantified by the dimensionless ratio
\begin{equation}
    \gamma \equiv \frac{V}{\kappa}.
\end{equation}

For the system considered in this work, we use  an approximate representation of the quantum spin state in the form of a matrix product state (MPS). This technique is a well-established approach to simulate the ground state and the dynamics of closed one-dimensional quantum many-body systems under well-controlled approximations~\cite{Vidal2004, Verstraete2008, Schollwock2011, Paeckel2019},  such as the dimension of the matrices or the time step (see Appendix~\ref{App: EstErr}). In order to simulate the master equation Eq.~\eqref{eq:mastereq}, we combine a time-dependent MPS algorithm (TEBD)~\cite{Paeckel2019} with a Monte-Carlo wave function method~\cite{Dalibard1992, Carmichael1993, Molmer1993}.  It is worth noting that, lately, a method based on MPS and a time-dependent variational principle was developed to simulate time evolution of many-body  quantum  lattice systems under continuous and local measurements~\cite{Doggen2021}.

For all numerical calculations (unless otherwise specified), we initialize the system in a Dicke state $\ket{D_N^{(k=N/4)}}$, and time-evolve it to generate a sufficiently large number of trajectories $\mathcal{M}$ ($\approx 250$). The observables are computed for each trajectory after the evolution has reached a steady-state, $\kappa t \gg 1$ (see Appendix~\ref{App: Steady-state} for details).

\subsection{State dependent observables}
In characterizing the trajectory dynamics of our system, we will consider several complementary measures.

\paragraph{Von Neumann entropy.}
The first quantity we study is the von Neumann entanglement entropy, which provides for pure states a measure of the entanglement built up between two parts of the system. Let us consider a chain of $N$ spins, which we divide into two subsystems A and B containing $\ell$ and $N-\ell$ sites, respectively. The entanglement entropy $S(\ell)$ is given by
\begin{equation}
    S_{\textrm{vN}}(\ell) \equiv S_{N}(\rho_{\ell} ) \equiv - \textrm{tr} (\rho_{\ell}  \textrm{log}_2 \rho_{\ell} ),
\end{equation}
where $\rho_\ell$ is the conditional reduced density matrix of the subsystem $\ell$. 
Entanglement is a fundamental property of quantum systems, widely used to characterize their critical properties and locate quantum phase transitions. In particular, the scaling of the entanglement with $\ell$ ($N$) in the asymptotic limit $\ell \to \infty$ ($N \to \infty$) gives us information about the nature of the phase. For instance, an entropy saturating with $\ell$ ($N$) indicates that the entanglement is proportional only to the surface of the block $\ell$ and no long-range correlation are built between the two subsystems. A logarithmic growth $S_{N}(\rho_L) \sim \log_{2}(\ell)$, instead, is characteristic of critical gapless phases.

\paragraph{Correlation functions.}
In addition to the entanglement, we investigate two additional indicators, the single-particle correlation function and its square. For instance, the quantum mechanical off-diagonal long-range order can be witnessed by the single-particle correlation functions
\begin{equation}
\label{eq: single_correlation}
     O_{i j} = |\langle \sigma_i^{+} \sigma_j^{-} \rangle| \neq 0,\: \text{for} \: |i-j| \rightarrow \infty.
\end{equation}
In contrast, short-range order is characterized by an exponential decay with the distance between sites.  Importantly, the single-particle correlation function averaged over an ensemble of trajectories is independent of the specificity of individual ones. As an alternative measure, we thus propose to study a related nonlinear quantity, namely the averaged square of the single-particle correlation functions:
\begin{equation}
\label{eq: square_correlation}
D_{ij} = \overline{|O_{i j}|^{2}} = \overline{|\langle \sigma_i^{+} \sigma_j^{-} \rangle|^{2}}.
\end{equation}

\section{ Entanglement entropy scaling transition}
\label{sec2}

\subsection{Entanglement entropy in the Dicke state}
\label{sec2.1}

\begin{figure}[t!]
    \centering
    \includegraphics[width=1\columnwidth]{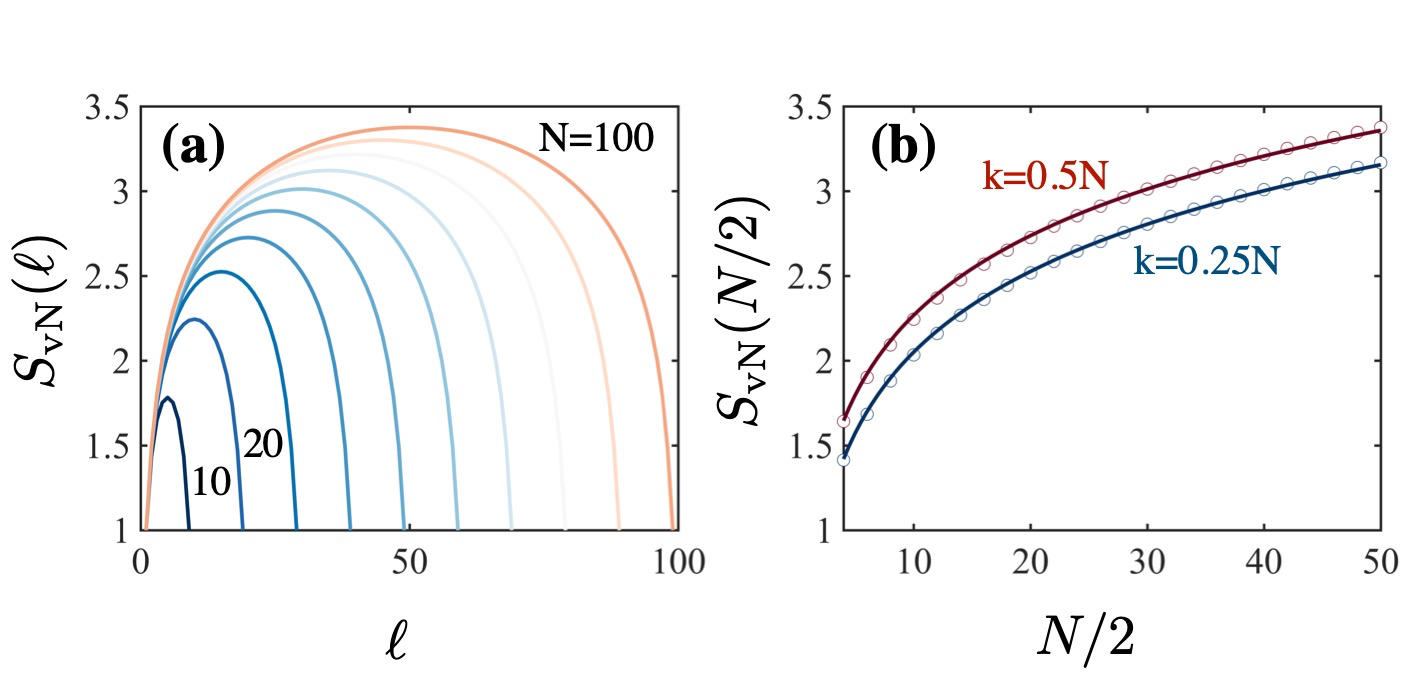}
    \caption{(a) Von Neumann entropy (Eq.~\eqref{eq: EE_dicke}) as a function of the block length $\ell$ and for system sizes N=10, ...100. (b) Half-chain von Neumann entropy, see Eq.~\eqref{eq: EE_dicke} and Eq.~\eqref{eq: max_S}, as a function of the system size $N/2$ for two different densities of excitations (see the color code). In both cases, the continuous lines represent the fit $S_{\rm vN} \approx a \log_2(N/2) + b$, with $a\approx 0.47 \pm 0.0036 $ and $b \approx 0.71 \pm 0.016$ for the red curve and $a\approx 0.48 \pm 0.0027 $ and $b \approx 0.47 \pm 0.011
    $ for the blue curve.}
    \label{fig: DS}
\end{figure}

 \begin{figure*}[t!]
  \centering
  \includegraphics[width=1\textwidth]{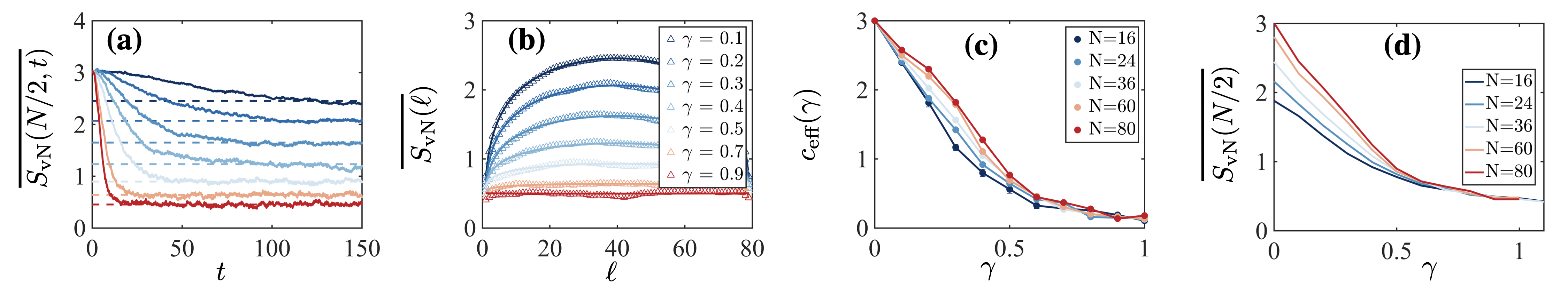}
  \caption{ (a) The time evolution of the averaged half-chain entropy initialized from $\ket{D_N^{(k=0.25N)}}$ for a chain of $N= 80$ spins for different $\gamma$. (b) Entropy for a chain of $N=80$ spins as function of the block length $\ell$. Continuous lines are fits of the form Eq.~\ref{eq: h_CC}. (c) The effective central charge $c_{\textrm{eff}}(\gamma)$ interpolated via a fitting procedure with Eq.~\ref{eq: h_CC} for different $N$ (see the color code) as function of $\gamma$. (d) Half-chain entropy for different chain of spins with a size $N=16, ... 80$ , as a function of the competition ratio $\gamma$.}
  \label{fig: Entropy}
\end{figure*}

We start by considering the extreme case $\gamma=0$. The steady-state solution corresponds to the fully coherent state, the Dicke state~\cite{Dicke1954}, see Eq.~\eqref{eq: Dicke_state}. In~\cite{Moreno2018}, the entanglement entropy was derived for all bipartitions on arbitrary Dicke states. The formula for a chain of $N$ spins corresponding to any bipartition $(\ell | N-\ell)$ and an arbitrary number of excitations $k$ is 
\begin{equation}
\label{eq: EE_dicke}
\begin{split}
S(N, k, \ell) = -\sum_{q=q'}^{q''} \frac{(N-\ell)!(N-k)! q!}{(N-\ell-k+q)! N!} \binom{\ell}{k}  \binom{k}{q} \\
\log_2 \Big[  \frac{(N-\ell)!(N-k)! q!}{(N-\ell-k+q)! N!} \binom{\ell}{k}  \binom{k}{q}  \Big], 
\end{split}
\end{equation}
where $q' = \textrm{max}(0, \ell+k-N)$ and $q'' =  \textrm{min}(\ell, k) $.

The bipartite EE thus grows  logarithmically with the system size. Furthermore, one can infer a generic upper bound for the entropy. Indeed, the function $S(N, k, \ell) $ is invariant under the permutation $\ell \leftrightarrow k$. In addition, this function presents a maximum for $k=N/2$, which due to the symmetry $\ell-k$, also corresponds to a maximum for $\ell = N/2$ for $N$ even~\cite{Moreno2018} (same reasoning holds for $N$ odd). After few steps, the maximal value of the entropy reads

\begin{equation}
    \lim_{n\to \infty} S_{\rm max}  = \frac{1}{2} \log_2(N/2).
    \label{eq: max_S}
\end{equation}

The expression Eq.~\eqref{eq: max_S} is reminiscent of the scaling of entanglement entropy for ground states of 1D critical Hamiltonians with open boundary conditions as given by the conformal field result
\begin{equation}
\rm S_{\rm vN}(\ell)  = \frac{c_{\rm eff}(\gamma)}{6} \log_2\Big[ \frac{N}{\pi}  \sin \big( \frac{\pi \ell}{N} \big) \Big] + s_0(\gamma).
\label{eq: h_CC}
\end{equation}
Here,  $c_{\rm eff}$ acts as an effective central charge, which is not given by an underlying conformal field theory (CFT),  and $\ell$ is the bipartite block length. Note that Dicke states are fully symmetric under the permutation group, restricting the subspace where the ground state must belong to. As we see from Eq.~\eqref{eq: max_S}, it leads to an upper bound, which implies that the entropy cannot grow faster than a logarithmic scaling~\cite{Latorre2009, Moreno2018}. This was shown in studies of the entanglement entropy in the Lipkin-Meshkov-Glick model~\cite{Latorre2005, Schachenmayer2013}.

Here, we use the effective central charge as an order parameter for the scaling of the entropy.
By using Eq.~\eqref{eq: max_S} and Eq.~\eqref{eq: h_CC}, we immediately obtain $c(\gamma =0) = 3$. Example results confirming these statements are plotted in Fig.~\ref{fig: DS}.

\subsection{Entanglement entropy as function of the interaction strength}
\label{sec2.2}

 Now, we focus on the averaged entanglement entropy for different ratios $\gamma$. The main result of our work is summarized in Fig.~\ref{fig: fig1} (c), where we compute the half-chain von Neumann entropy $\overline{S_{\textrm{vN}}(N/2)}$ for different values of the competition ratio $\gamma$. In Fig.~\ref{fig: fig1} (c), continuous lines are fits obtained by using the scaling form Eq.~\eqref{eq: h_CC}. We identify three regimes: (i) in the absence of competition ($\gamma=0$), the steady-state corresponds to the Dicke state with long-range off diagonal order (black line). We have shown in Sec.~\ref{sec2.1}  that the bipartite entropy for the Dicke state features a logarithmic scaling. (ii) We find a logarithmic dependence of the entanglement entropy for small competition ratio, similar to a conformal field theory with open boundary conditions. (iii) For large competition, the entropy is constant with the system size $N$, which indicates a disentangled area law phase. In Fig.~\ref{fig: fig1} (d), we report the effective central charge $c_{\textrm{eff}}(\gamma)$ and the residual entropy $s_0(\gamma)$ (inset), extracted by fitting the data with Eq.~(\ref{eq: h_CC}), as a function of $\gamma$. We see that $c_{\textrm{eff}}(\gamma)$ is finite for $\gamma < \gamma_c$, while for $\gamma >\gamma_c$, $c_{\textrm{eff}}(\gamma)=0$. This result etablishes an area law transition at non-zero ratio $\gamma_c\approx 0.5$. 
 
 In Fig.~\ref{fig: Entropy} (a), we compute the time evolution of the  half-chain entropy averaged over a few hundred trajectories ($\mathcal{M} \approx 250$). The entropy reaches a plateau indicating its steady-state value. The dashed lines represent the mean steady-state values and are a guide for the eye. In Fig.~\ref{fig: Entropy} (b) we show the entanglement entropy as a function of the block length $\ell$ for various values of $\gamma$. We find that for small $\gamma$, the entropy scales logarithmically with $\ell$, while for large $\gamma$, the entropy is constant for all $\ell$. By using the functional form Eq.~(\ref{eq: h_CC}), we fit our data  (continuous lines) in order to obtain the parameters $c_{\textrm{eff}}(\gamma)$ and $s_0(\gamma)$. The parameter $c_{\textrm{eff}}(\gamma)$ is shown in Fig.~\ref{fig: Entropy} (c) for different system sizes. We infer a critical point $\gamma_c\approx [0.4-0.6]$, which is in good agreement with the numerical results obtained in Fig.~\ref{fig: fig1} (d). Finally, in Fig.~\ref{fig: Entropy} (d), we show the steady-state scaling of the half-chain entropy as a function of $\gamma$ to observe how the behavior changes across the critical point $\gamma_c$ (dotted lines). For $\gamma < \gamma_c$, the curves for different system sizes (see the color code) split up. In contrast, for $\gamma > \gamma_c$, these curves overlap, confirming that the half-chain entropy is independent of the size of the chain $L$, in agreement with the behavior observed in Fig.~\ref{fig: fig1} (c).
 
\section{Correlation functions} 
\label{sec3}

\begin{figure*}[t!]
  \centering
  \includegraphics[width=1\textwidth]{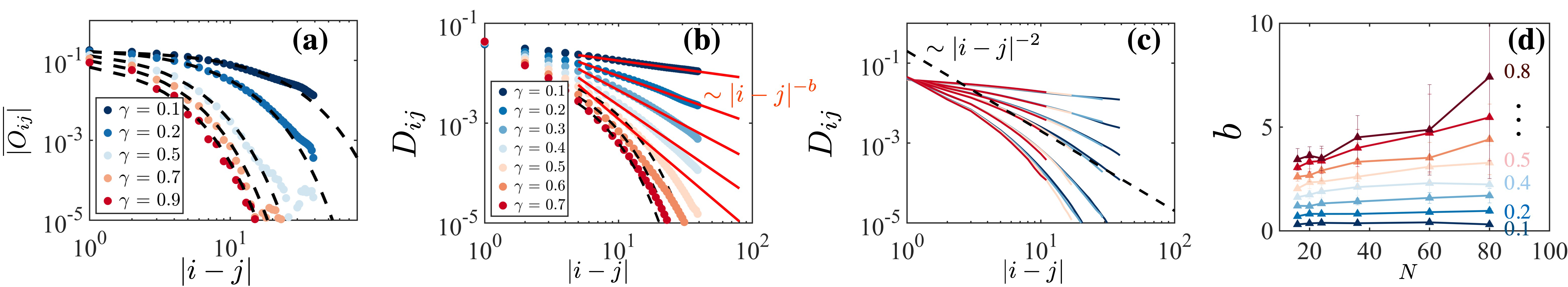}
  \caption{(a) Averaged two-point correlator  $\overline{|\langle \sigma_i^{+} \sigma_j^{-} \rangle|}$ as function of the distance $|i-j|$ for different interaction strengths. (b) Averaged square of the single-particle correlation functions $D_{ij}$ for various $\gamma$. Red lines are best fits of the form $a/x^{b}$. Black lines are heuristic fits of the form $ b e^{(- a |i-j|)}$ and reflect the short-range order. (c) $D_{ij}$ for different system sizes  $N = 20, 24, 36, 60, 80$. It shows a data collapse. (d) Exponent of the algebraic decay as a function of the system size $N$, for different $\gamma$ (see the color code).}
  \label{fig: obdm}
\end{figure*}

In this section, we study how the correlation functions decay with increasing ratio $\gamma$. We compute the single-particle correlation functions, see Eq.~\eqref{eq: single_correlation}, $O_{i_0 j_f}$ from site $i_0= L/4$ to site $j_f= L-L/4$ at maximum in order to minimize the finite-size effects. We average them over $256$ realizations and different times. We have checked that the correlation functions have reached their steady-state value for each time considered (see Appendix ~\ref{App: Steady-state}). 

In Fig.~\ref{fig: obdm} (a), we find that the averaged one-body density matrix decays exponentially for any finite $V$, i.e. $\overline{O_{i j}} \simeq \textrm{exp}(-|i-j|)$. The presence of any finite interaction ($V\neq 0$) destroys off-diagonal long-range order indicating that at a mean steady-state  dynamics the systems ends up irremediably in a short-range order phase. 

In Fig.~\ref{fig: obdm} (b), we show $D_{ij}$ as a function of the distance $|i-j|$ for different $\gamma$ (see color code) and $L=80$. In stark contrast, with the averaged single-particle correlation functions, at small $\gamma < \gamma_c$, we observe an algebraic decay of the correlation functions $D_{ij}$. The red lines are best fits of the form $a/x^{b}$. When crossing the transition to the area law regime $\gamma > \gamma_c$, the correlations start to decay more rapidly with the distance $|i-j|$ between different sites. At large distance $|i-j| \gg 1$, a heuristic fit $D_{ij} \approx \textrm{exp}(-|i-j|)$ (dotted black lines) reflects the short-ranged correlations.
 
The data collapse of the correlation functions for different system sizes is shown in Fig.~\ref{fig: obdm} (c). In Fig.~\ref{fig: obdm} (d), we show the exponent of the algebraic decay of the correlation functions. To extract the exponent $b$,  we suppose that the correlations decay as a power law, $D_{ij} = a/|i-j|^{b}$ . Then, we extract the mean exponent $b$ for different system sizes by taking its logarithmic derivative. If the correlations decay algebraically, the exponent remains constant for each system size. In Fig.~\ref{fig: obdm} (d), we present the mean exponent $b$ (the errorbar corresponds to the standard deviation of the mean) as a function of the system size $N$ and for various $\gamma$, ranging from $0.1$ (lower plot) to $0.8$ (upper plot). We see that the exponent stays relatively constant up to a $\gamma \approx 0.5$. However, for $\gamma > 0.5$, the exponent fluctuates strongly with the system sizes $N$, which indicates that the correlations do not decay algebraically anymore. 

Note that the error bars on the correlation functions (not shown for clarity) are large. In particular, at large distance $\Delta_{O_{i j}} \approx 10^{-2}$ and $\Delta_{D_{i j}} \approx 10^{-3}$. This effect is explained by the statistical error intrinsic of the quantum trajectory algorithm and the error of the MPS representation (see Appendix~\ref{App: EstErr} for more details). 


\section{A quantum trajectory analysis}
\label{sec4}

\begin{figure*}[t!]
  \centering
  \includegraphics[width=0.8\textwidth]{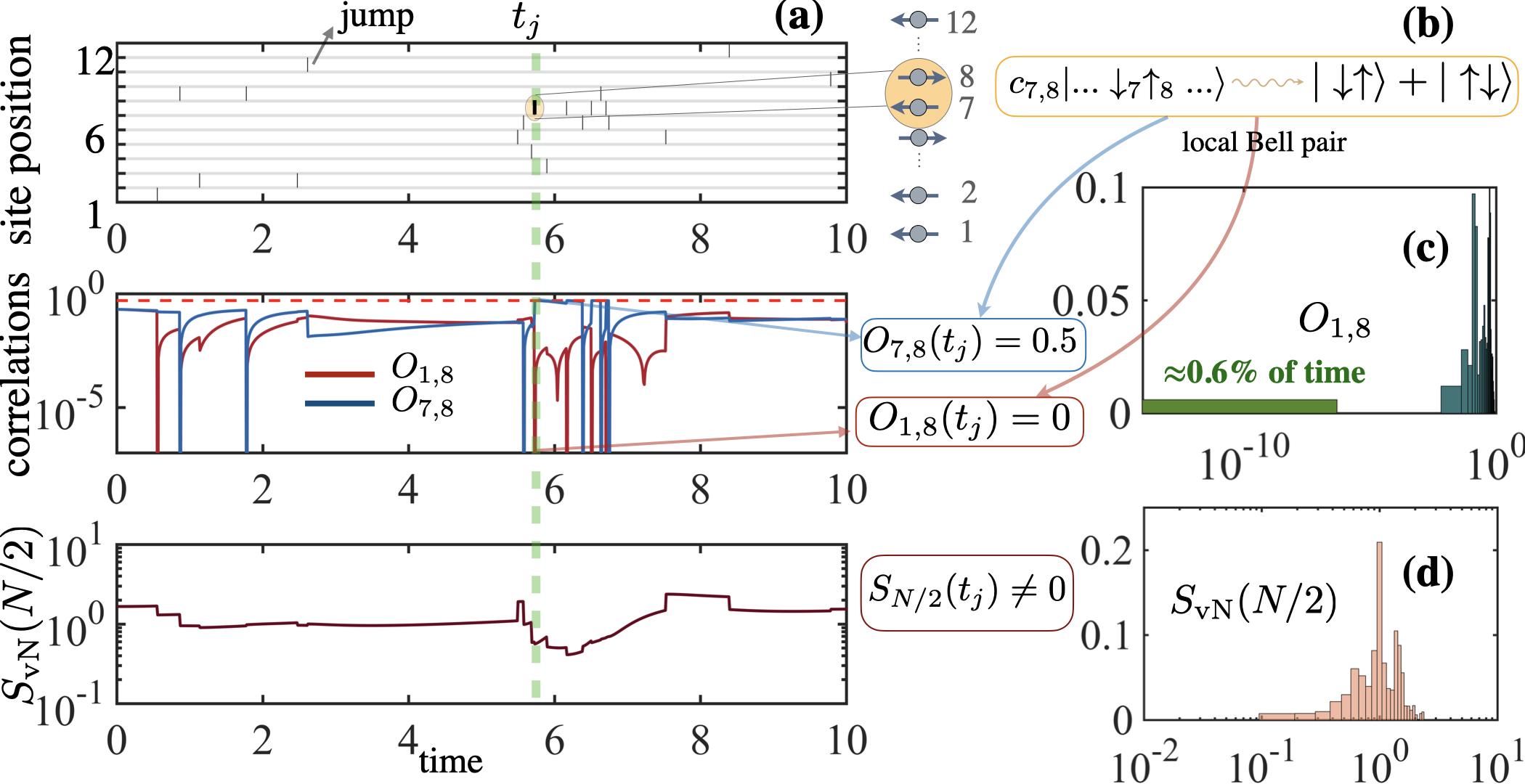}
  \caption{(a) Upper panel: position in space ($y$-axis) and time ($x$-axis) of quantum jump events (black line) during one trajectory. Middle panel: time evolution during a trajectory of the single-particle correlation $O_{1 8}$ (red line) and $O_{7 8}$ (blue line). The red dashed line represents the maximum of the correlation. Lower panel: time evolution of the half-chain von Neumann entropy during a trajectory. The vertical green dashed line is an illustrative example. It corresponds to the time $t_j$ where a jump occurs between sites 7 and 8. (b) Schematic view of the action of the jump operator (here at time $t_j$ for sites 7 and 8). (c) Distribution of the correlation value $O_{1 8}$ during one trajectory. The green bar represents the zero value, set to [$10^{-12},10^{-8}$] for illustration purpose only. (d) Distribution of $S_{\rm v N}(N/2)$ during a single trajectory. In the distribution (c) and (d), the $x$-axis represents the value of the quantities considered and the $y$-axis their probability of occurence. The parameters underlying the simulation are $\gamma=0.5$, $\textrm{d}t=0.01$, $N=12$.}
 \label{fig: traj}
\end{figure*}

Hitherto, we have focused on averaged quantities, but further information on the dynamics can be inferred from the study of observables along a single trajectory. In particular, in this section we study the evolution of selected observables during a single trajectory, such as the single-particle correlation function and the EE, and analyse how the occurrence of quantum jump events affects them. We show that jump operators can annihilate certain correlations while the entropy remains less affected, which we trace back to the local action of our engineered dissipative term (Eq.~\eqref{eq:jumpope}). Then, we show that the MIT can be found by studying the distribution of the entanglement entropy over the duration of a single trajectory. We find that the log-to-area law transition is characterized by a uni-to-bi modal distribution of the entropy. Additionally, we show that the distribution of the entropy along a single trajectory is equivalent to an average over many trajectories. By doing so, we evidence a self-averaging property of the EE, where the single trajectory EE (averaged over a sufficiently long time) behaves like the EE averaged over an ensemble of independent trajectories.

We begin by studying how a jump operator affects our system and selected observables. In Fig.~\ref{fig: traj} (a), we compute the time evolution of the single-particle correlation functions $O_{1 8}$, $O_{7 8}$ (middle panel), and the half-chain entropy (lower panel) for $\gamma = 0.5$, $N=12$, and a time step $\textrm{d}t=0.01$. The upper panel represents the position in space and time of the quantum jump events. Then, let us consider a time $t_j$ immediately after a jump has occurred between sites 7 and 8 (see green dashed vertical line). This two-site operator creates locally a  maximally entangled Bell pair, as illustrated in panel (b). While the local correlation is maximal ($O_{7 8}=0.5$, see the blue line), due to the monogamy of entanglement, spins 7 and 8 completely decouple from all other spins of the system, which leads to a complete suppression of all other correlators involving site 7 or 8, as shown with $O_{1 8} (t_j)=0$ (red line). This phenomenon leads to large fluctuations in the correlation functions. In Fig.~\ref{fig: traj} (c), we compute the distribution of the correlation Eq~\eqref{eq: single_correlation} during a single trajectory ($t_{\rm max} = 50$). We observe that the correlation spends a non-negligible time at strictly zero value (green bar, set to [$10^{-12},10^{-8}$] for illustration purposes only), leading to an asymmetrical weight distribution. In stark contrast, the half chain entanglement entropy is not very sensitive to jump events. Its value does not fluctuate much during one trajectory, as shown in the lower panel of Fig.~\ref{fig: traj} (a). In Fig.~\ref{fig: traj} (d), we show the entanglement entropy distribution.  

\begin{figure}[h!]
  \centering
  \includegraphics[width=1.0\columnwidth]{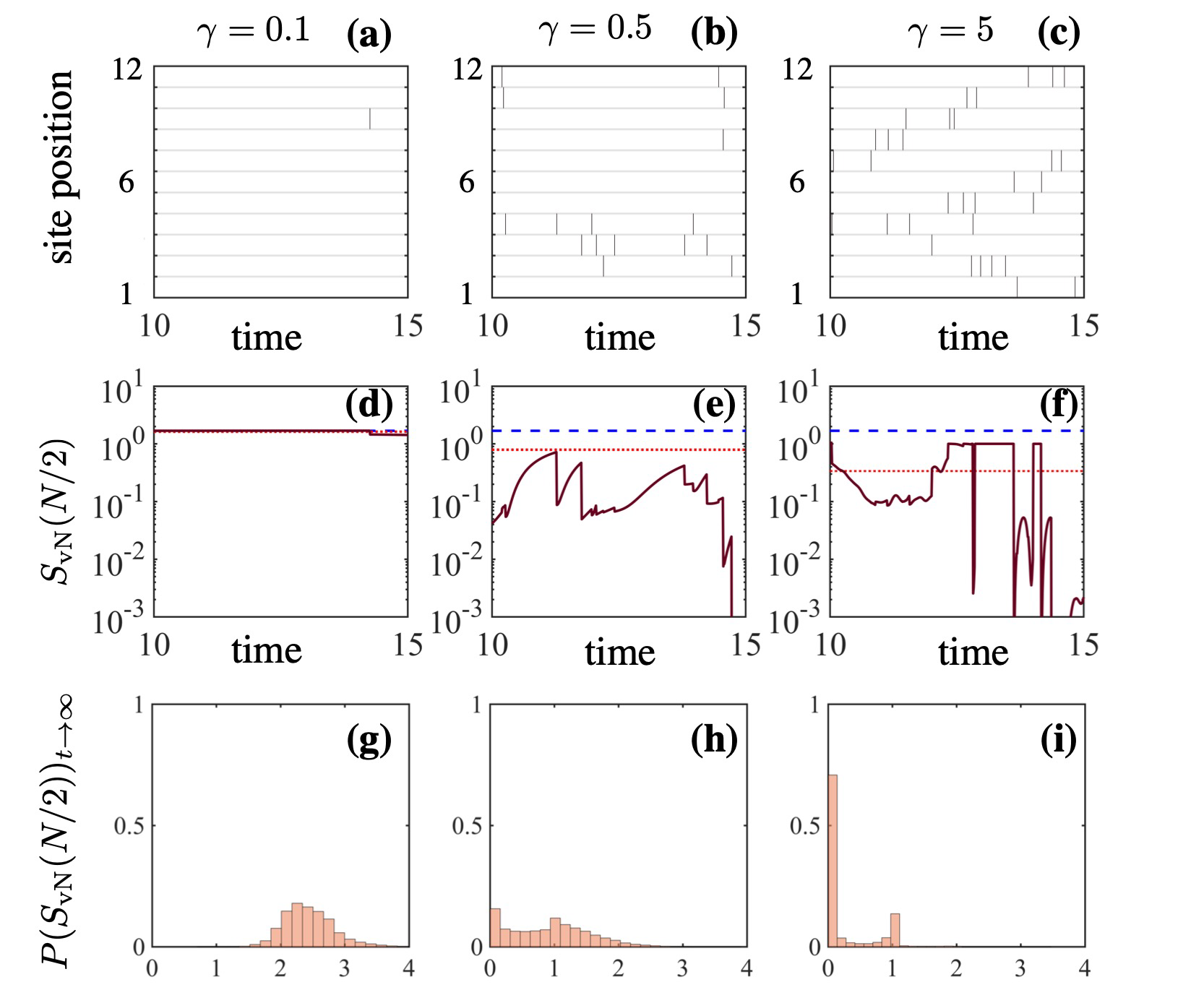}
  \caption{Upper panels (a)-(c): position in space ($y$-axis) and time  ($x$-axis) of quantum jump events (black line) during one trajectory. Middle panels (d)-(f): time evolution during a trajectory of the half-chain von Neumann entropy. Lower panels (g)-(i): distribution of $S_{\rm v N}(N/2)$  averaged over $256$ trajectories. In panel (d)-(f), the red dotted lines are the mean value of the entropy during this trajectory and the blue  dashed line is the exact value of the entropy for the Dicke state. The parameters are timestep $\textrm{d}t=0.01$, competition ratio $\gamma=0.1$ (left panels), $\gamma=0.5$ (middle panels) and $\gamma=5$ (right panels), and system size $N=12$ for panels (a)-(f), and $N=60$ for panels (g)-(i).}
 \label{fig: E_distri}
\end{figure}

Focusing on the entropy, we find that the fluctuations in its temporal distribution along a single long enough trajectory reveals the scaling transition. In Fig.~\ref{fig: E_distri}, we indicate the position and time of jump operators acting on a chain of 12 spins [(a)-(c)], and the corresponding bi-partite half chain entropy [(d)-(f)]. The panels [(g)-(i)] illustrate, for a larger size  $N=60$, the distribution of $S_{\rm vN} (N/2)$ averaged over $256$ trajectories.  Note that, when we average the distribution of the entropy in each trajectory, we remove the time below 10 to avoid possible effects due to the initial equilibration dynamics after initialisation in the Dicke state.  Importantly, we also find that the distribution of the entropy of a single long trajectory is equivalent to the averaged distribution of the entropy. The test has been realized for a system size of $N=12$ sites and results are shown in the Appendix~\ref{App: add_num_result}.

We present three cases, $\gamma=0.1$ in the left panels, $\gamma=0.5$ in the middle panels, and $\gamma=5$ in the right panels. The red dotted lines correspond to the mean value of the entropy. The blue lines are the exact value of the bi-partite entropy for the Dicke state $\ket{D_N^{(k=0.25N)}}$ obtained from Eq.~\eqref{eq: EE_dicke}. At small strength of the competing Hamiltonian potential [$\gamma=0.1, $ Fig.~\ref{fig: E_distri} (g)], we find that the distribution of the half chain entropy ($P(S_{\rm vN}$)) is unimodal and normally distributed. In this case, the entropy deviates slightly from its mean value.  Close to the critical ratio [$\gamma_c=0.5, $ Fig.~\ref{fig: E_distri} (h)], however,  the entropy starts to become asymmetrically distributed with two peaks located at zero and one. We have checked that this characteristic signature persists for larger values of $\gamma$, as illustrated in panel (i) for $\gamma=5$. The bimodality of the distribution stems from the fact that some trajectories are entangled with a value of the EE close to the one of the Dicke state, while others are reduced to zero (weakly) entangled states, and the system exhibits temporal fluctuations between the two extremes, as is visible e.g.~in Fig.~\ref{fig: E_distri} (f).

\begin{figure}[t!]
  \centering
  \includegraphics[width=1.0\columnwidth]{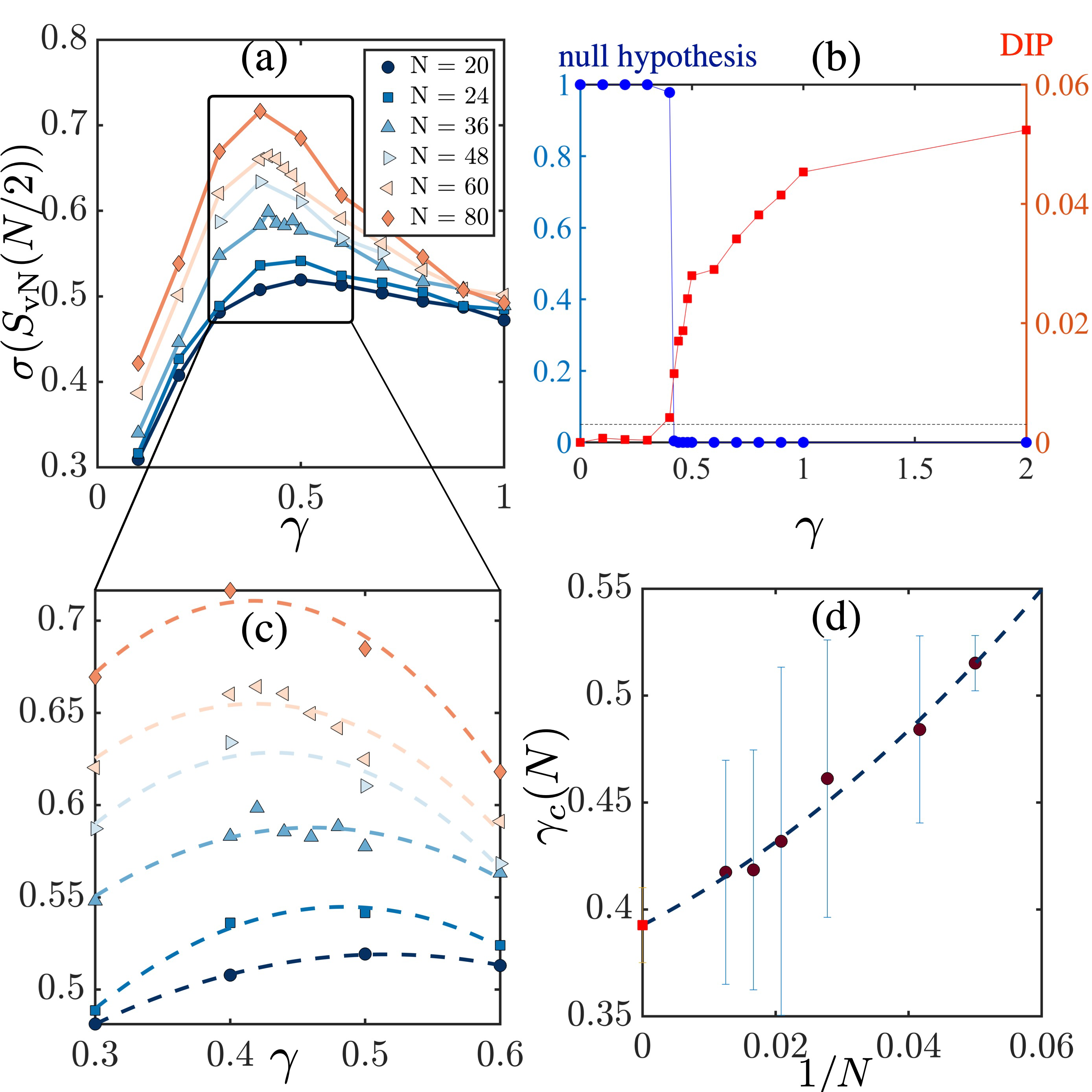}
  \caption{(a) Standard deviation of the half chain entanglement entropy ($\sigma_{S_{\textrm{vN}}(N/2)}$) as a function of $\gamma$, for different system sizes (see color code). (b) Hartigan's dip test (red) and null hypothesis test (blue) as a function of $\gamma$, for a system size $N=60$. The dotted line is the probability estimate (result: $0.99$) for the data to be described by a distribution with more than one mode (see Appendix~\ref{App: dip} for details). (c) Zoom-in of the $\sigma_{S_{\textrm{vN}}(N/2)}$ in the region around the peak of the distribution. Dashed lines are fits of the form $ax^{2} + b x + c$, from which we extrapolate $\gamma_{c}(N)$ corresponding to a maximum of $\sigma_{S_{\textrm{vN}}(N/2)}$. (d) Finite size analysis of $\gamma_{c}(N)$. The dashed line is a heuristic fit  of the form $a/N^{2} + b/N + c$. The critical point (red dot) $\gamma_c (N\to \infty) \approx 0.39 \pm 0.02$.}
 \label{fig: dip_test}
\end{figure}

Statistically, various tests exist to detect the presence of more than one mode in a distribution. Here, we propose to use Hartigan's dip test~\cite{Hartigan1985}. In this approach, we test the null hypothesis of unimodality by realizing the dip test of unimodality  that determines the probability of an empirical distribution function being bimodal. A large value of the dip indicates that the distribution of analysed data is more probable to have multiple modes. By combining this method with an analysis of the standard deviation, we determine the critical point of the transition. Results of Hartigan's dip test are shown in Fig.~\ref{fig: dip_test} (b) for $N=60$. The null hypothesis and the dip test cross at $\gamma= [0.4-0.5]$. In Fig.~\ref{fig: dip_test} (a), we find that the standard deviation of  $S_{\rm vN} (N/2)$ ($\sigma_{S_{\textrm{vN}}(N/2)}$) has a peak located at $\gamma= [0.4-0.5]$. In order to obtain a more precise estimate, we extrapolate $\gamma_{c}(N)$ corresponding to the maximum of $\sigma_{S_{\textrm{vN}}(N/2)}$ using a fit of the form $ax^{2} + b x + c$ close to the location of the peak. Examples are illustrated in Fig.~\ref{fig: dip_test} (c), where dotted lines are the fits. Using these extrapolated values, we then realize a finite-size analysis of $\gamma_{c}(N)$ as shown in Fig.~\ref{fig: dip_test} (d). The dotted line is a heuristic fit of the form $a/(N^{2}) + b/N + c$. We find a critical point $\gamma_c (N\to \infty) \approx 0.39 \pm 0.02$. These results are in agreement with the critical value obtained from the entanglement entropy analysis. We note that the observed bimodality is not an artifact of the system sizes considered, or the trajectories. For more detailed results see Appendix~\ref{App: add_num_result}. 

We note that this structural change (bimodality) of the distribution of the entanglement entropy was already observed in~\cite{Alberton2021}. Complementarily, we have emphasized new features by showing that (i) this bimodality property is present at a single (long) trajectory level. This approach opens the possibility that this type of scaling transition can be highlighted by studying the entropy propagation for long enough times along single trajectories. (ii) Remarkably, moderate system sizes of about only 30 spins are sufficient to clearly establish and quantitatively analyse the transition by means of this feature. This renders this tool into an attractive and efficient alternative to the numerically more demanding analysis of the scaling of the EE for system sizes of 80-100 spins.

\section{Interacting spin model}
\label{sec5}

\begin{figure*}[t!]
  \centering
  \includegraphics[width=1\textwidth]{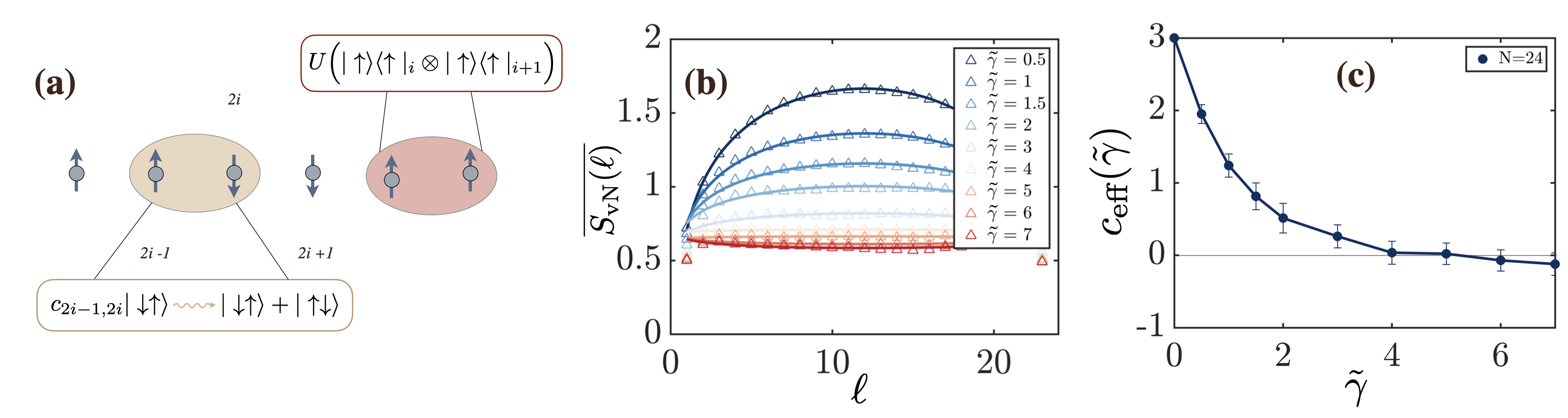}
  \caption{(a) A chain of spins with nearest-neighbors interaction and jump operators of the form Eq.~\eqref{eq:jumpope}. (b) Von Neumann entropy as a function of the block lenght $\ell$ for a system size $N=24$. Lines are fits of the form Eq.~\eqref{eq: h_CC}. (c) Effective central charge extracted from the scaling form fit Eq.~\eqref{eq: h_CC} as a function of the competition ratio $\tilde{\gamma}$. Transition is found at $\tilde{\gamma}_c \leq 4$}
 \label{fig: model2}
\end{figure*}

We now show that the discussed scaling transition is found for a broader class of Hamiltonians competing with the engineered dissipation. For concreteness, we consider 
\begin{equation}
\label{eq: second_model}
    H  = \frac{U}{4} \sum_i^{N}  (1 +  \sigma^{z})_i (1 + \sigma^{z})_{i+1},
\end{equation}
where $U$ denotes the strength of the interaction potential. As in the previous case, we define a dimensionless ratio
$ \tilde{\gamma} \equiv U/\kappa$ ($\kappa \equiv 1$).
This Hamiltonian describes interaction of spin excitations (or hard-core bosons) located on neighbouring sites, as illustrated in Fig.~\ref{fig: model2} (a). This coherent dynamics has similar effects to the staggered potential studied previously. The pure dark state (Dicke state), which contains spatially delocalised excitations, and which is stabilised by the engineered dissipation, is not an eigenstate of the Hamiltonian~\eqref{eq: second_model}. Therefore, the Hamiltonian will lead to dephasing and a destruction of long-range order, thereby also reducing the entanglement in the system. Therefore, we expect that the competition between this coherent and the dissipative dynamics leads to a MIT. Note that the impact of the Hamiltonian~\eqref{eq: second_model} depends on the density of spin excitations since only nearest-neighbors spin excitations will interact. Here, we focus on a fixed density and leave the analysis for various excitation densities  for future study. In the following, we apply the same approach and parameters as used in the previous model. 

Fig.~\ref{fig: model2} (b) shows typical results of $S_{\rm vN}(\ell)$ as a function of the block length $\ell$ for different interaction strengths $U$. Lines are fits of the form Eq.~\eqref{eq: h_CC}. We find that the entropy scales logarithmically with $\ell$ for small values $\tilde{\gamma}$, while for large $\tilde{\gamma}$ it becomes constant as a function of $\ell$. We conclude that a transition between a critical regime with a logarithmic growth of the entanglement and an area law regime is also present in this model.
By using the scaling form in Eq.~\eqref{eq: h_CC}, we can extract an effective central charge $c_{\textrm{eff}}(\tilde{\gamma})$ for different values of $\tilde{\gamma}$. Taking into account the error on the fit, our numerical results suggest a transition around $\tilde{\gamma}\leq 4$. Other observables, such as the distribution of the entropy or correlation functions are consistent with this result -- see Appendix~\ref{App: add_num_result} for a more detailed discussion, corresponding plots and numerical values.

\section{Discussion}
\label{sec: Discussion}

In this paper, we have demonstrated the presence of a measurement-induced transition in a so far unexplored and complementary scenario, where the dissipation stabilizes a logarithmic growth of the EE,  while the Hamiltonian induces an area law scaling. We have shown that the transition can manifest itself in other state-dependent observables, such as correlation functions nonlinear in the state. 
Additionally, we have found that the distribution of the entropy of a single trajectory constitutes an efficient indicator of the transition. This self-averaging property is manifest also by the emergence of bimodality in its distribution, even for small to moderate system sizes. This is particularly relevant to highlight MITs in more generic interacting models where numerical simulations are a challenging endeavour. Thus, understanding the robustness of such criterion in other scenarios and different MITs could lead to an efficient diagnostic tool that overcomes the usually more demanding numerical analysis of the entanglement entropy or correlation functions. More broadly,  our work could pave the way towards addressing the question of observing different MITs in driven many-body systems. For instance, it could be potentially applied to explore a situation where the non-unitary dynamics generates a chaotic behavior in the search for a volume-to-area law transition. In light of our findings, it would also be interesting to develop an analytical theory for the models presented in this work and in similar ones, as it was recently done, e.g., for Dirac fermions in one dimension with continuous measurements~\cite{Buchhold2021}.

\begin{acknowledgments}
We are grateful to Johannes Schachenmayer for stimulating discussions and for help with the MPS code. TB and MM acknowledge support by the European Research Council (ERC) via ERC Starting Grant QNets Grant Number 804247. MM acknowledges support by the EU H2020-FETFLAG-2018-03 under Grant Agreement number 820495. SD acknowledges support by the  Deutsche Forschungsgemeinschaft (DFG, German Research Foundation) under Germany’s Excellence Strategy Cluster of Excellence Matter and Light for Quantum Computing (ML4Q) EXC 2004/1 390534769, by the DFG Collaborative Research Center (CRC) 183 Project No. 277101999 - project B02, and by the European Research Council (ERC) under the Horizon 2020 research and innovation program, Grant Agreement No. 647434 (DOQS).  We thank the Centre de calcul de l'Universit\'e de Strasbourg where we carried out most of the numerical investigations.
\end{acknowledgments}

\appendix
\section{Overlap with the Dicke State}
\label{App: ConvDickeState}
\begin{figure}[h!]
    \centering
    \includegraphics[width=1\columnwidth]{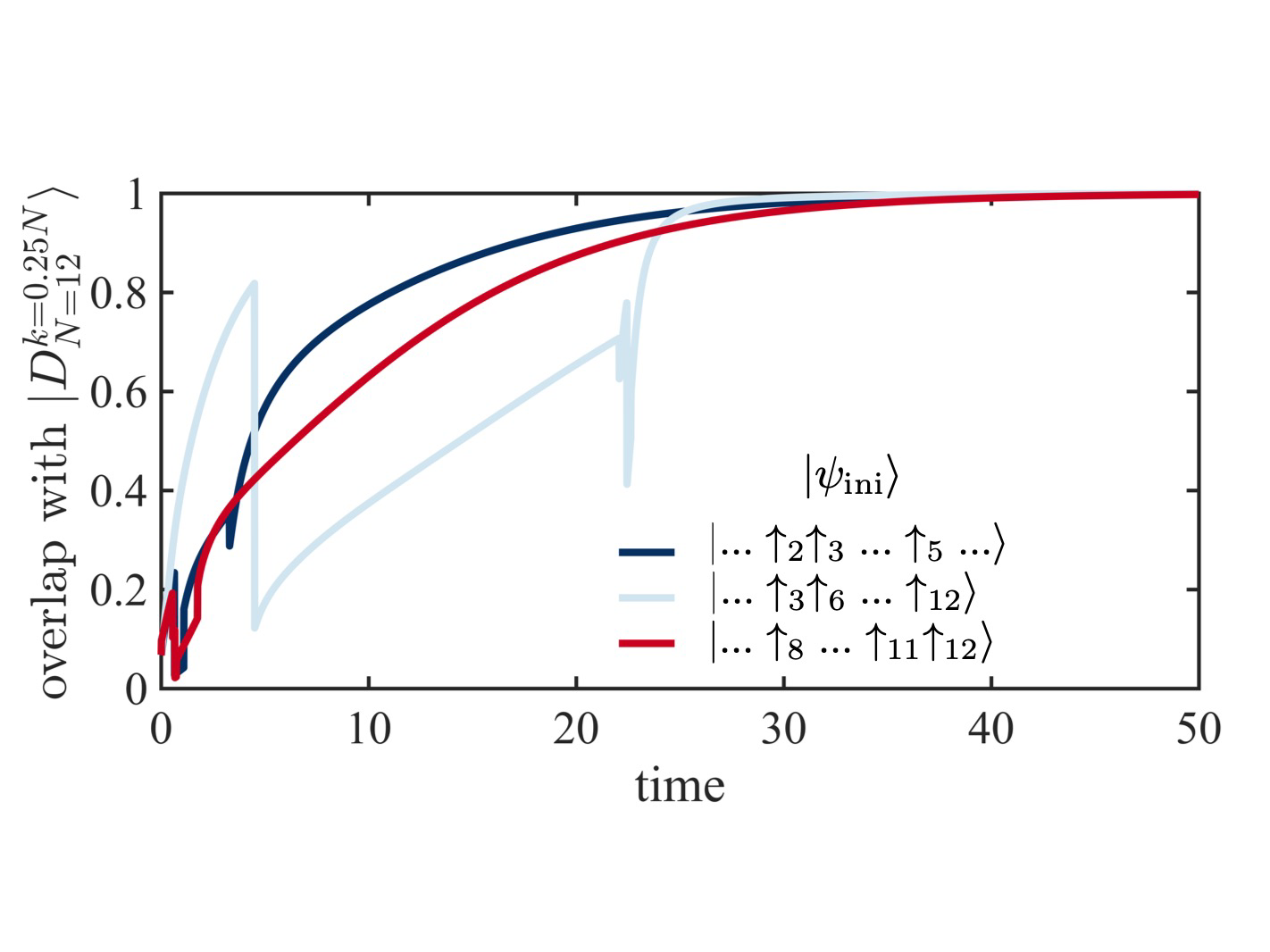}
    \caption{In a chain of $N=12$ spins, time evolution of the overlap between different initial states, each containing $k = N/4$ spin excitations (see color code), and the Dicke State $|D_{N=12}^{k=3} \rangle$.
    }
    \label{fig: fig1_supp_mat}
\end{figure}

In this section, we consider the case $V=0$, corresponding to purely dissipative dynamics. We check that independently of the initial conditions, the system ends up in the Dicke state (see Eq.~\eqref{eq: Dicke_state}). In Fig.~\ref{fig: fig1_supp_mat}, we show the temporal evolution of the overlap between several initial states and the Dicke state $|D_{N=12}^{k=3}\rangle$ during a single trajectory up to a time $t=50$. We choose different random initial states (see caption), all with $k=N/4$ excitations. They all converge, after a certain time, to the Dicke State $|D_{N=12}^{k=3}\rangle$.

\section{Steady-state}
\label{App: Steady-state}
\begin{figure}[t!]
    \centering
    \includegraphics[width=1\columnwidth]{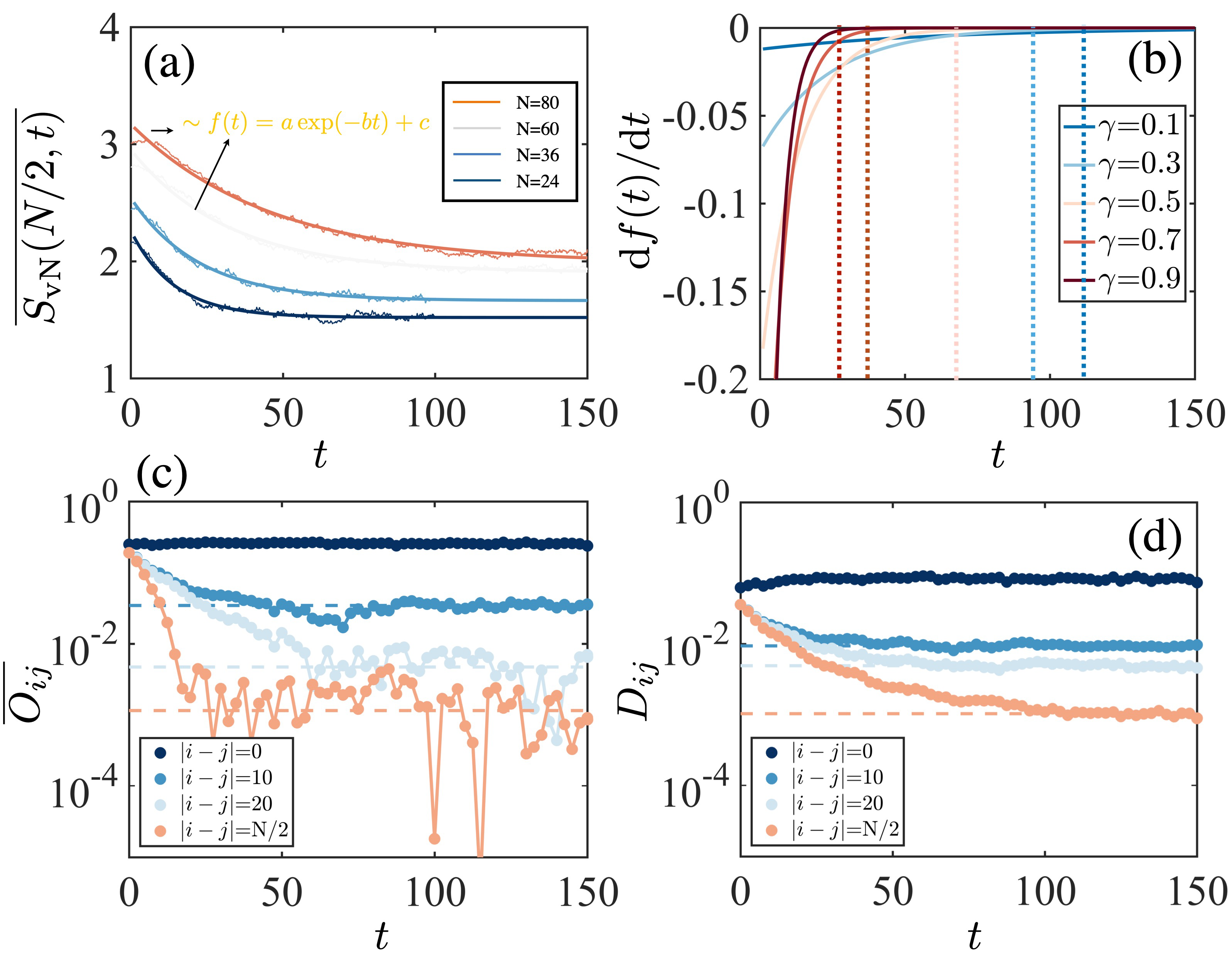}
    \caption{(a) Time evolution of the half chain von Neumann entropy averaged over 256 trajectories for $\gamma=0.2$ and different sizes $N=24, 36, 60, 80$. Lines are fits of the form  $f(t) = a\exp(-b t) + c$. (b) Derivative of $f(t)$ ($\dv{f(t)}{t}$) as a function of the time $t$ for different $\gamma$ and $N=80$. (c) Single-particle correlation function $\overline{O_{ij}}$ and (d) its square $D_{ij}$ for different distances $|i-j|$. The dashed lines are the mean steady-state value and are a guide to the eye.}
    \label{fig: fig2_supp_mat}
\end{figure}

In this section we discuss how we determine the steady-state time $t_s$ for the entanglement entropy and the correlation functions. 
In Fig.~\ref{fig: fig2_supp_mat} (a), we present the half-chain entropy averaged over 256 trajectories as a function of the time for $\gamma =0.2$. The lines are fits of the form $f(t) = a\exp(-b t) + c$. By taking the derivative of the previous functional form, we can  approximately infer the value $t_s$, which corresponds to $\dv{f(t)}{t} \approx 0 $, as illustrated in Fig.~\ref{fig: fig2_supp_mat} (b). To be sure we have reached the steady state, we have always considered times $t\geq t_s$. In Fig.~\ref{fig: fig2_supp_mat} (c) and (d), we compute the single-particle correlation function and its square averaged over 256 trajectories  as a function of time for $\gamma=0.2$ and different distances $|i-j|$. We used $i_0=N/4$ for the initial site and $j_{\rm f}= 3N/4$ for the last one to reduce the finite size effect. Despite significant fluctuations (we comment about that in Sec.~\ref{App: EstErr}), we see that correlation functions converge to a plateau after a certain time. In order to improve the statistical average we obtain the final average expectation value of an observable $\hat{O}$ using: 
\begin{equation}
\label{eq: sum_t}
\overline{\hat{O}} =  \frac{1}{\mathcal{N}} \sum_{t_s}^{t_{\rm max}} \overline{\hat{O_t}}, 
\end{equation}
where $\mathcal{N}$ is the number of $t \geq t_s$. Note that we have saved the value for the half chain entropy at each time step, while for the correlation functions, we have saved them every 250 steps.

\section{Estimate of the numerical errors}
\label{App: EstErr}

\begin{table*}
\centering
\begin{tabular}{|c|c|c|c|c|c|c|}
 \multicolumn{7}{|c|}{Table of numerical parameters for $\gamma = 0.1$} \\
  \hline
  \hline
  N (system size) & dt (time step) & tmax & BD  & $\sum_t \epsilon(t)$ (total truncation error) & $\mathcal{M}$ (\# of trajectories) & final error entropy\\
  \hline
  16 & 0.01 & 100 & 30 & $10^{-12}$ & 256 & \textemdash{} \\
  \hline
   20 & 0.01 & 100 & 30 & $10^{-7}$ & 256 &\textemdash{} \\
  \hline
  24 & 0.01 & 100 & 30 & $10^{-6}$ & 256 & \textemdash{} \\ 
    \hline
   36 & 0.01 & 100 & 50 & $10^{-5}$ & 256 &  $\approx 10^{-2}$ \\
    \hline
  48 & 0.01 & 100 & 60 & $3\times10^{-5}$ & 256 &\textemdash{} \\
  \hline
    60 & 0.01 & 150 & 80 & $1\times10^{-5}$ & 256 &\textemdash{} \\
  \hline
  80 & 0.01 & 150 & 100 & $1\times10^{-4}$ & 256 &  $\approx 10^{-2}$\\
  \hline
\end{tabular}
\caption{Table that summarizes the numerical parameters used during the simulation for $\gamma = 0.1$}
\label{table: table_numerics}
\end{table*}

In this section we add some additional details regarding the estimate of the numerical errors. We consider two type of errors in our systems. On the one hand, there is the statistical error inherent from the wavefunction method~\cite{Daley2014}. For any operator $\hat{0}$ this errors is estimated as
\begin{equation}
\label{eq: stat_err}
\sigma_0 = \frac{\Delta_0}{\sqrt{\mathcal{M}}}, 
\end{equation}
where  $\Delta_0$ is the standard deviation (or sample estimate) of $\overline{\hat{0}}$ and $\mathcal{M}$ is the number of sample (corresponding to the total number of trajectories).
On the other hand, we must consider the error due the MPS time evolution approach we used (TEBD).  Specifically, TEBD suffers from two types of errors. Firstly, the time step error, which is of order $\Theta(\Delta_t^3)$, with $\Delta_t$ the time step, for a second-order Trotter decomposition. In all calculations, we have chosen $\Delta_t=0.01$ leading to an error of $\Theta(\Delta_t^3) = 10^{-6}$. Secondly, a realistic representation of a quantum state as a MPS requires a truncation of the Hilbert space, which is controlled by the `bond dimension'' (BD). For a finite BD, the precision of the results is determined by the convergence with the BD, as well as the truncation error $\epsilon$ (sum of discarded squared singular values) during each optimization process. In Table.~\ref{table: table_numerics}, we show the parameters underlying the simulation for $\gamma=0.1$. Similar parameters have been used for larger $\gamma$.

\subsection{Entropy}
\begin{figure}[t!]
    \centering
    \includegraphics[width=1\columnwidth]{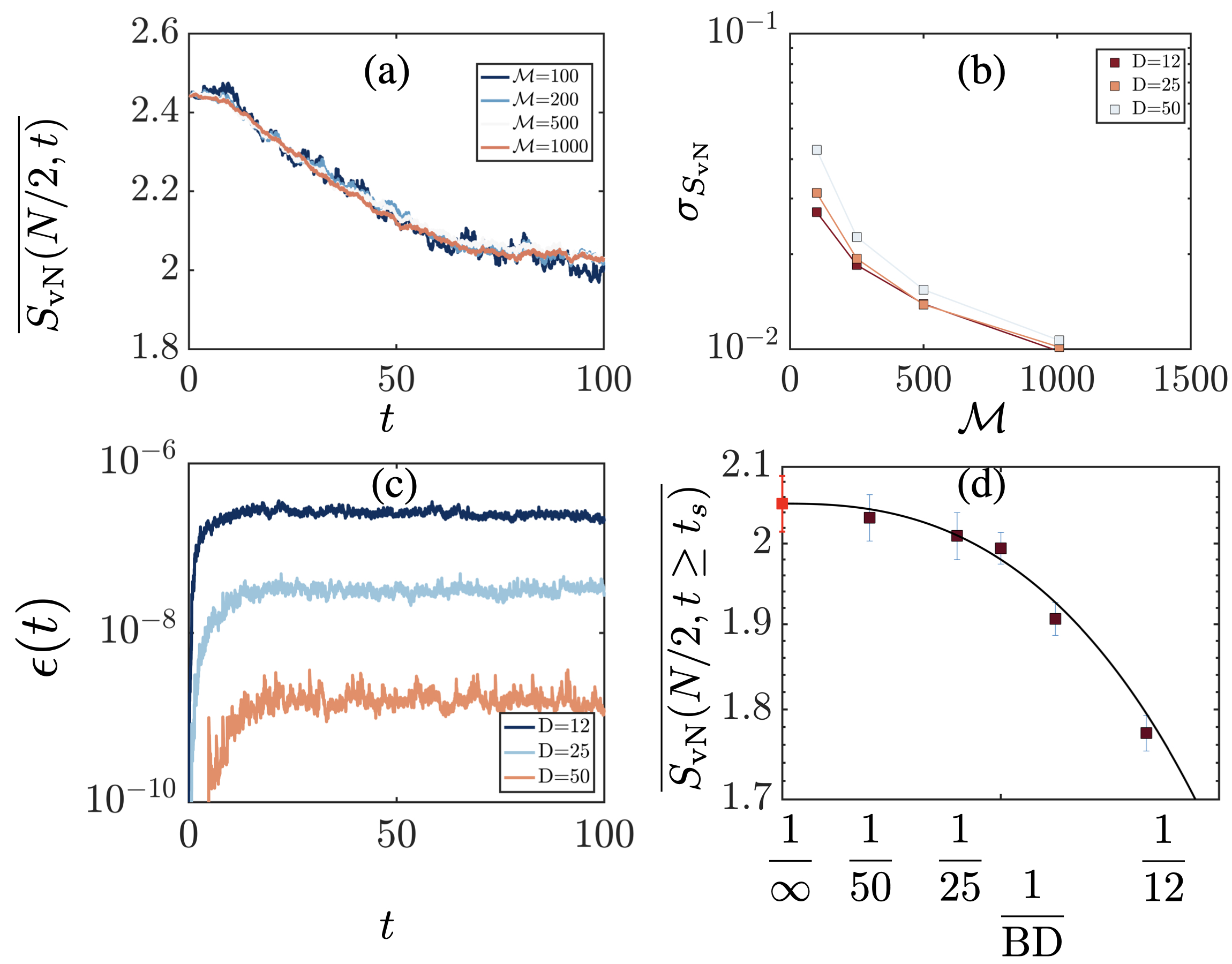}
    \caption{(a) Time evolution of the half chain von Neumann entropy averaged over different number of trajectories $\mathcal{M} = 100, 250, 500, 1000$ for $N=36$. (b) Statistical error $\sigma_{S_{\rm vN}}$ as a function of  $\mathcal{M}$. (c) Sum of the truncation error occurring at each time step as a function of the time for different BD$=12, 25, 50$. (d) Finite-BD scaling of $\overline{S_{\rm vN}(t \geq t_s)}$. The error bars originate from the statistical error $\sigma_{S_{\rm vN}}$ in panel (b). The line is a fit of the form $a/x^b + c$. The red dot corresponds to the extrapolated value  $S_{\rm vN} (\text{BD} \to \infty)$.}
    \label{fig: fig3_supp_mat}
\end{figure}

We first focus on the statistical error $\sigma_{S_{\rm vN}}$ of the entanglement entropy. In Fig.~\ref{fig: fig3_supp_mat} (a), we show the time evolution of the entropy for different numbers of trajectories $\mathcal{M}$ and a BD $=50$. Using Eq.~\eqref{eq: stat_err}, we extract the statistical error as a function of $\mathcal{M}$. The error is of order  $0.9 \times 10^{-1}$ for the number of  trajectories considered in the main text ($\approx 250$). However, we have used Eq.~\eqref{eq: sum_t} in order to improve the average. We thus consider the value  $10^{-2}$ as the upper limit of the statistical error. 
Then we determine the error related to the bond dimension. In Fig.~\ref{fig: fig3_supp_mat} (d) we show an example of extrapolating $S_{\rm vN}$ in the limit of infinite bond dimension for $\gamma=0.1$. The line is a heuristic fit of the form $a/x^b + c$. Here, the error bars come from the estimate $\sigma_{S_{\rm vN}}$ obtained in Fig.~\ref{fig: fig3_supp_mat} (b). We finally obtained the error by considering the difference $|S_{\rm vN} (BD \to \infty) -S_{\rm vN} (BD =50)| \approx 2\times 10^{-2}$. The standard errors in the fit parameters are obtained from the least square method. The latter are included in our fitting procedure and thereby in the final error. It is also important to note that a truncation error arises when we apply a gate operator (MPO). In particular, during one-time step of the TEBD algorithm, we subsequently apply two site gate operators on the whole chain. In order to keep track of this error, we sum all the truncation errors resulting from the application of these gates during a time step, it reads
\begin{equation}
\epsilon(t) = \sum_{i=1}^{N-1} \epsilon_{i, i+1} + \sum_{i=N}^{i+1} \epsilon_{i-1, i}, 
\end{equation}
where $\epsilon_{i, i+1}$ is the truncation error resulting from the application of a two-site gate operator on site $i$ and $i+1$. $\epsilon(t)$ is displayed in Fig.~\ref{fig: fig3_supp_mat} (c) for different BD. Finally, by summing $\epsilon(t)$ one can see that the total error due to the truncation increases with time. The latter gives a good estimate of the total error we get for a particular bond dimension.

\begin{figure}[t!]
    \centering
    \includegraphics[width=0.99\columnwidth]{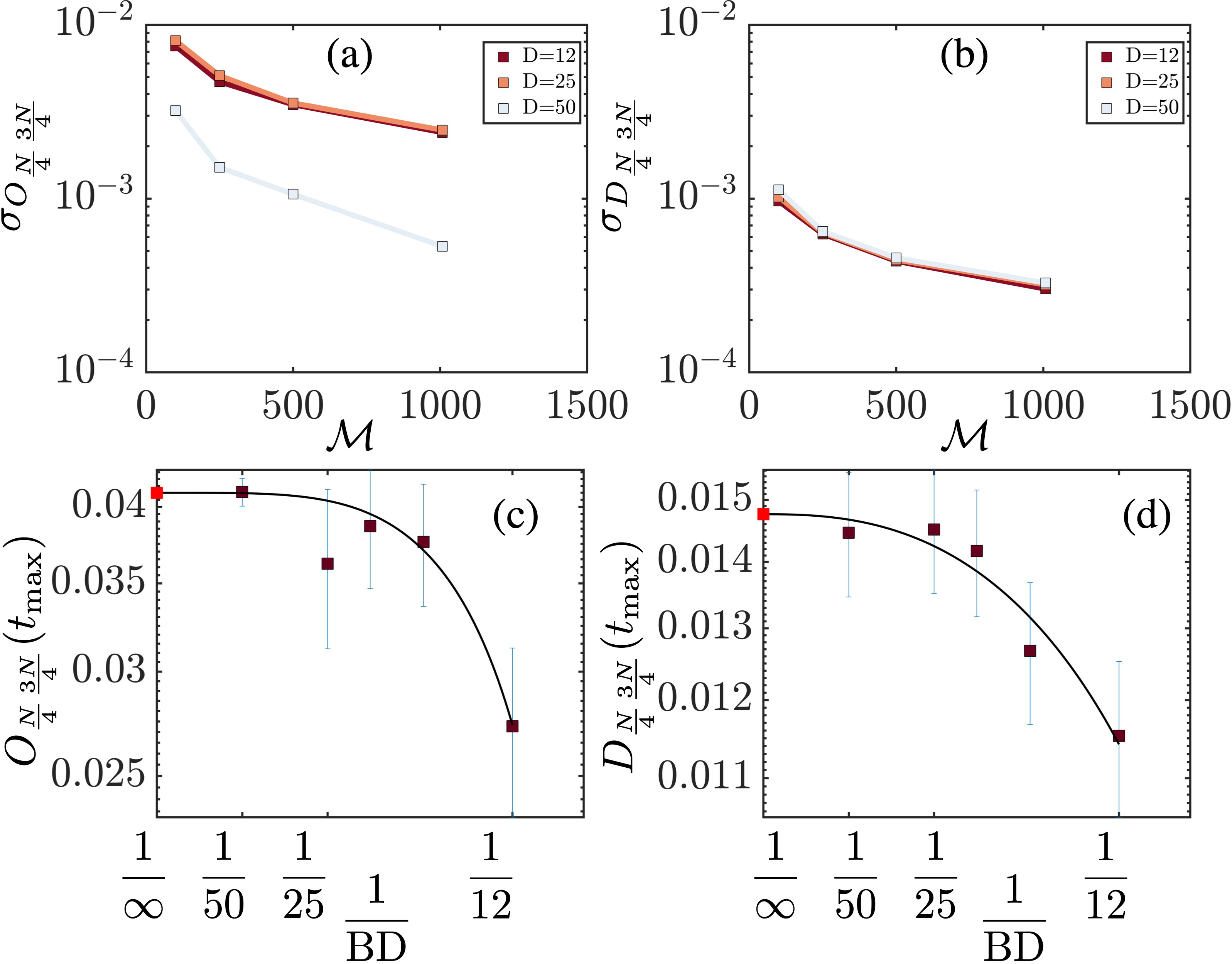}
    \caption{(a) , (b) Statistical error $\sigma_{O_{ij}}$ and $\sigma_{D_{ij}}$ as a function of  $\mathcal{M}$ for the maximum distance considered in the main text, i.e. $i_0=N/4$ and $j_{\rm f}= 3N/4$. (c), (d) Finite-BD scaling for $\overline{O_{ij}}$ and $D_{ij}$, respectively. The lines are fits of the form $a/x^b + c$. Red dots correspond to the extrapolated values. The data are for $N=36$.}
    \label{fig: fig4_supp_mat}
\end{figure}

\subsection{Correlation functions}

We now estimate the uncertainties in the correlation function $\overline{O_{i_0 j_{\rm f}}}$ (Eq.~\eqref{eq: single_correlation}) and $D_{{i_0 j_{\rm f}}}$ (Eq.~\eqref{eq: square_correlation}) where $i_0=N/4$ and $j_{\rm f}= 3N/4$. This situation corresponds to the maximum distance considered in the main text. The statistical errors are displayed in Fig.~\ref{fig: fig4_supp_mat} (a), (b) for $\overline{O_{i_0 j_{\rm f}}}$ and $D_{{i_0 j_{\rm f}}}$, respectively. In Fig.~\ref{fig: fig4_supp_mat} (c), (d), we show $\overline{O_{i_0 j_{\rm f}}}$ and $D_{{i_0 j_{\rm f}}}$ as a function of the $1/{\rm BD}$. The error bars are the statistical errors obtained in Fig.~\ref{fig: fig4_supp_mat} (a), (b) for $\mathcal{M} = 250$. In order to extrapolate the correlators in the limit of infinite BD, we  fit our data with a functional  form $a/x^b + c$. Importantly, we note that the statistical errors on $\overline{O_{i_0 j_{\rm f}}}$ (or equivalently on  $D_{{i_0 j_{\rm f}}}$) are included in our fitting procedure. We thereby determine the final error by computing $|\overline{O_{i_0 j_{\rm f}}(\text{BD}\to \infty)} - \overline{O_{i_0 j_{\rm f}}(\text{BD} =50)}| \approx 10^{-2} $ and $|\overline{D_{i_0 j_{\rm f}}(\text{BD}\to \infty)} - \overline{D_{i_0 j_{\rm f}}(\text{BD} =50)}| \approx 3\times10^{-3}$. Here, we stress that the relative error on the different correlation functions becomes quickly important at large distance. This effect leads  to a significant uncertainty in the tail of the correlation functions displayed in Fig.~\ref{fig: obdm} (a), (b) and (c).

\section{Additional numerical results}
\label{App: add_num_result}

\subsection{First model (staggered potential)}

\subsubsection{Distribution of the entropy}

\begin{figure}[th!]
    \centering
    \includegraphics[width=1\columnwidth]{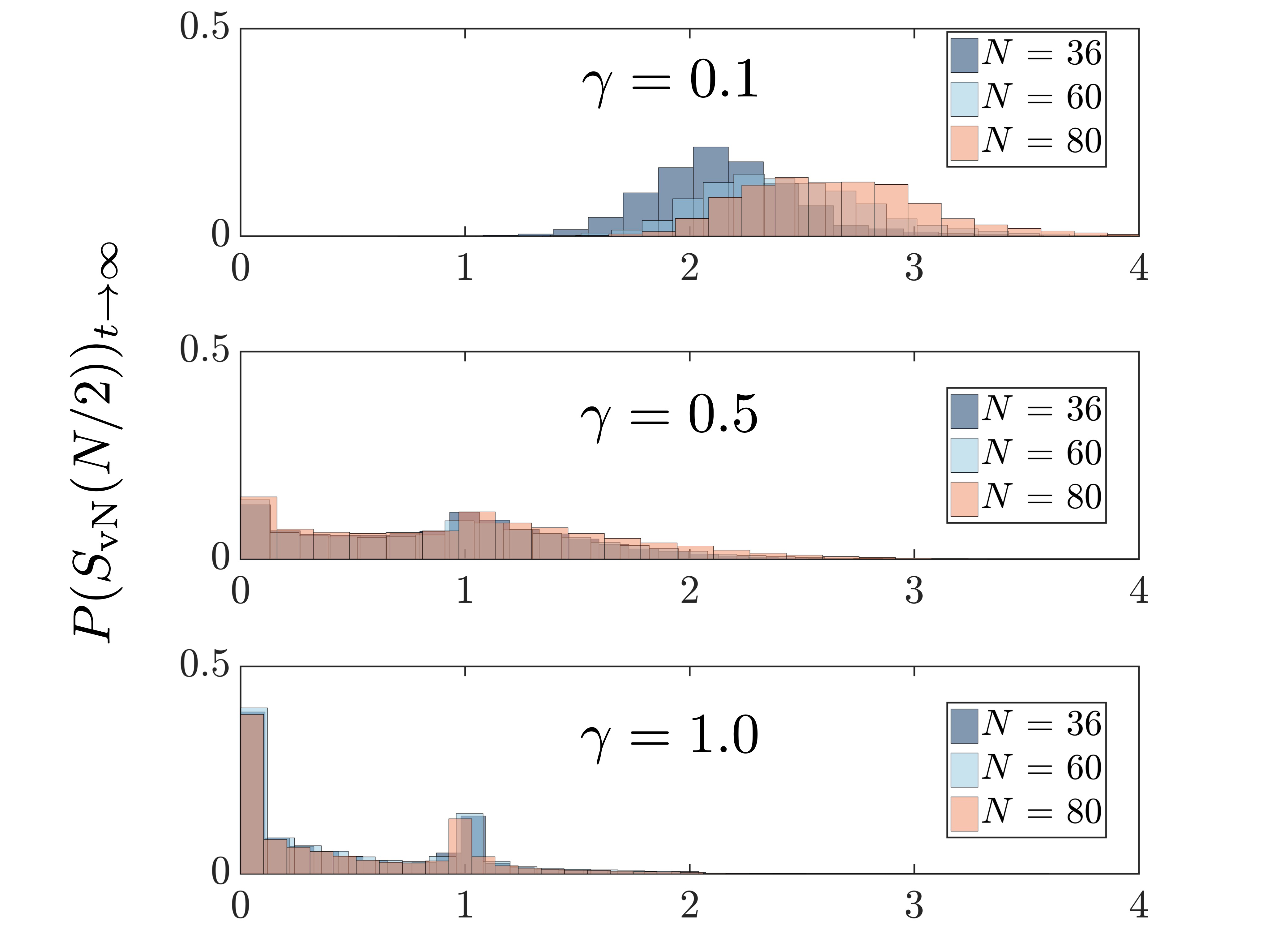}
    \caption{Distribution of $S_{\rm v N}(N/2)$ averaged over $\mathcal{M}=256$ trajectories and different system sizes $N=36, 60, 80$ (see the color code). Upper panel $\gamma = 0.1$, middle panel $\gamma =0.5$, lower panel $\gamma= 1$.}
    \label{fig: fig_distri_supp_mqt}
\end{figure}

In the main text, we have shown that the distribution of the half-chain entropy during a long trajectory reveals the transition. Indeed, while at small $\gamma$ the distribution of the entropy is unimodal, at large $\gamma$, we have observed a bimodal distribution. Here, we check that this particular behavior is not an artifact of finite size or specific realizations. In Fig.~\ref{fig: fig_distri_supp_mqt}, we show the distribution of the half-chain entropy averaged over $256$ trajectories for $\gamma=0.1, 0.5, 1.0$ and different system sizes (see the color code). We find that independently of the system size, the distribution is bimodal when we are in the regime $\gamma \geq \gamma_c \approx 0.5$. 

\subsubsection{Self-averaging property}

\begin{figure}[th!]
    \centering
    \includegraphics[width=1\columnwidth]{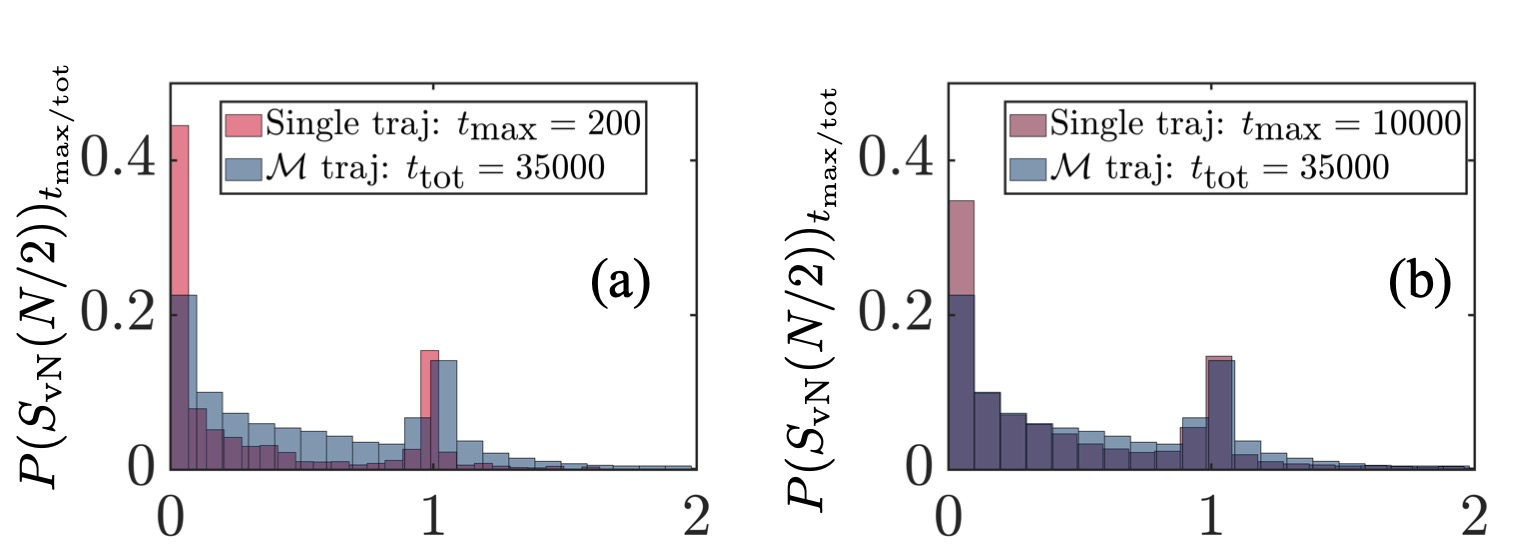}
    \caption{Distribution of $S_{\rm v N}(N/2)$ for $N=12$ as obtained from an average over $\mathcal{M}=256$ trajectories  ($t_{\rm tot} \approx 35 000$) and from a single long trajectory with $t_{\rm max} = 200$ (a) and $t_{\rm max} = 10 000$ (b).}
    \label{fig: self_averaging}
\end{figure}

In this section, we demonstrate the self-averaging property of the entropy. We compare the averaged distribution of the entropy over many independent trajectories ($\mathcal{M}=256$), and a single trajectory runs for two times $t = 200, 10000$. In Fig.~\ref{fig: self_averaging}, we show the distribution of the entropy for $N=12$ computed from the averaged of $\mathcal{M}$ trajectories and a single trajectory simulate up to a maximum time $t_{\textrm{max}}=200$ in panel (a) and $t_{\textrm{max}}=10000$ in panel (b). The distribution of the single trajectory converges to the averaged distribution when we increase the maximum time of its simulation. This indicates that both methods are equivalent. We always used the averaged approach since it is computationally more convenient.

\subsection{Second model (nearest-neighbors interaction) }
\begin{figure}[th!]
    \centering
    \includegraphics[width=1\columnwidth]{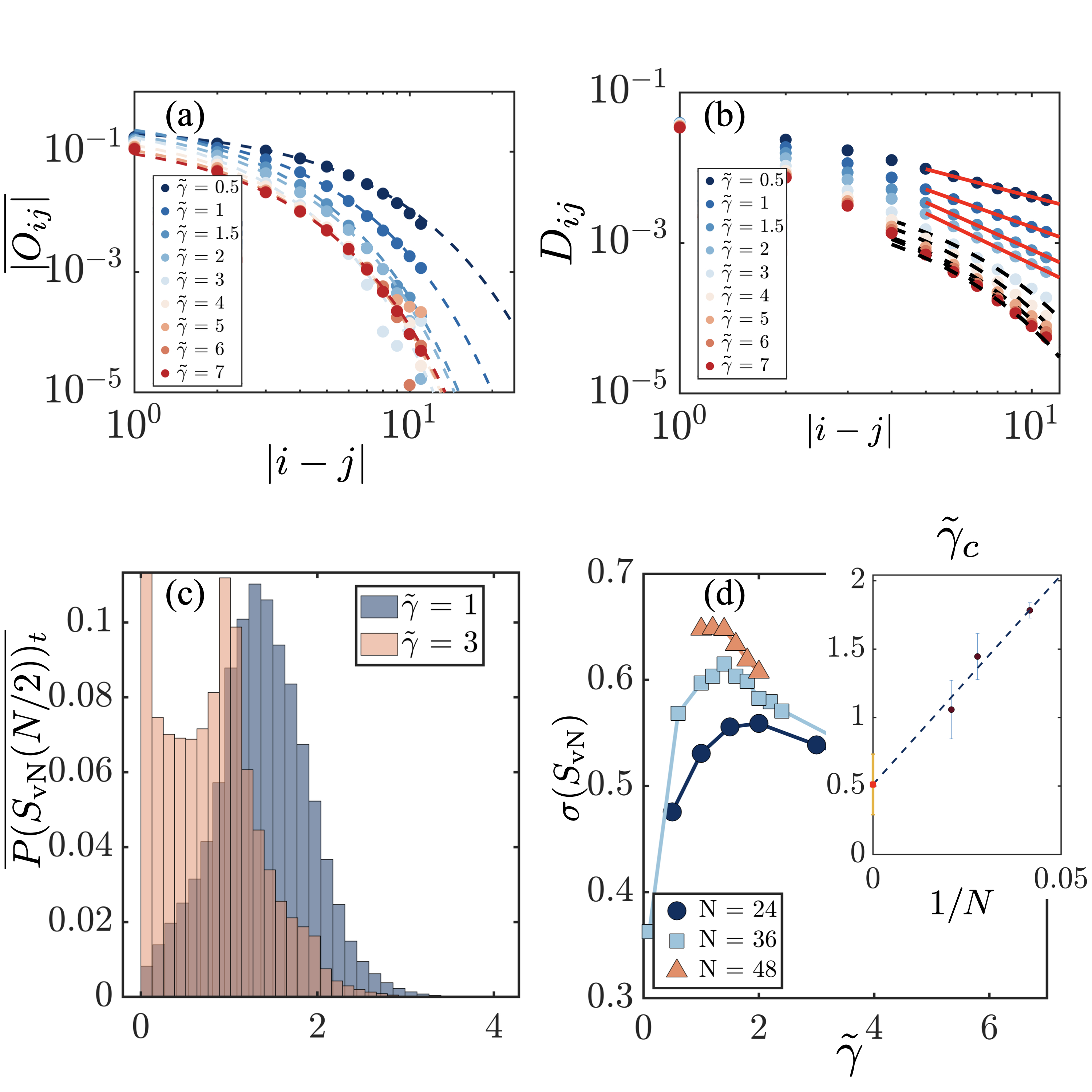}
    \caption{(a), (b) Single-particle correlation function $\overline{|O_{i j}|}$ and its square $D_{i j}$ as a function of the distance $|i-j|$ for different $\tilde{\gamma} = U/\kappa$ ($\kappa \equiv 1$). Dashed lines are heuristic fits of the form $a\exp(|i-j|)$ and red lines are fits of the form $a/|i-j|^{\alpha}$. (c) Distribution of the half chain entanglement entropy averaged from $\mathcal{M}=256$ trajectories for two interaction strengths $\tilde{\gamma}=1$ and $\tilde{\gamma}=3$. The $x$-axis is the value of $S_{{\rm vN}}$ and the $y$-axis is its probability of occurrence. (d) Standard deviation of the distribution of the entropy as a function of $\tilde{\gamma}$. The inset shows the finite size scaling of $\tilde{\gamma_c}$ corresponding to the maximum of the variance. The latter leads to an estimate of the critical point $\tilde{\gamma_c} \approx 0.5 \pm 0.22 $. These maxima are extrapolated from a fit of the form $a x^{2} + b x +c$ around the peak. All data are for $N=24$.}
    \label{fig: fig_second_model_supp_mat}
\end{figure}
In this section, we add a numerical complement for the second model we have studied in the main text (see Eq.~\eqref{eq:  second_model}). In Fig.~\ref{fig: fig_second_model_supp_mat} (a) and (b), we compute the single-particle correlation function $\overline{|O_{ij}}|$  and its square $D_{ij}$, respectively. We see that $\overline{|O_{ij}}|$ decays exponentially (dashed lines are heuristic exponential fits) as a function of the distance $|i-j|$. It indicates a short range order phase. In contrast, $D_{ij}$ scales as a power law for $\tilde{\gamma} < 2$ (red lines are fits of the form $a/|i-j|^{b}$). For $\tilde{\gamma} > 2$, $D_{ij}$ seems to decay exponentially (dashed black line). However, the system size considered here is small, thereby making a clear distinction between and exponential decay and an algebraic one is challenging. As a consequence, we can not precisely determine the critical point of the transition from Fig.~\ref{fig: fig_second_model_supp_mat} (b).

Then, we study the temporal distribution of the entropy averaged over $256$ trajectories for $\tilde{\gamma}=1$ and $\tilde{\gamma}=3$, as illustrated in Fig.~\ref{fig: fig_second_model_supp_mat} (c). We see that for $\tilde{\gamma}=1$ the distribution of the entropy is unimodal, while for $\tilde{\gamma}=3$ the distribution of the entropy is bimodal. This behaviour is reminiscent of the one observed in Fig.~\ref{fig: E_distri}. To estimate the critical value, we use the method develops in Sec.~\ref{sec4}. We extrapolate $\tilde{\gamma}_{c}(N)$ corresponding to the maximum of $\sigma_{S_{\textrm{vN}}}$ in Fig~\ref{fig: fig_second_model_supp_mat} (d) using a fit of the form $ax^{2} + b x + c$ close to the location of the peak. Then, we perform a finite-size analysis of $\tilde{\gamma}_{c}(N)$ as shown in the inset of Fig~\ref{fig: fig_second_model_supp_mat}. We have used a fit of the form $ a x + b$ (dashed line) and we obtain an estimate of the critical point  $\tilde{\gamma}_{c}(N) = 0.5 \pm 0.22$.

\section{Hartigan's dip test of unimodality}
\label{App: dip}

\begin{table}[b!]
    \centering
\begin{tabular}{ |p{1cm}||p{1cm}|p{1cm}|p{1cm}| p{1cm}| p{1cm}| p{1cm}|  }
\hline
 Sample size & \multicolumn{6}{|c|}{Probability of dip less than tabled value} \\
 \hline
  & 0.01  & 0.1 & 0.90 & 0.95 & 0.995 & 0.999 \\
 \hline
    4   & 0.1250 & 0.1250 & 0.1863 & 0.2056  & 0.2387 & 0.2458 \\
    5   & 0.1000 & 0.1000 & 0.1773 & 0.1872 & 0.1981 & 0.1996 \\
    6   & 0.0833 & 0.0833  & 0.1586 & 0.1645 & 0.2034 & 0.2224 \\
    20  & 0.0474 & 0.0569 & 0.0970 & 0.1047 & 0.1262 & 0.1382 \\
    50  & 0.0312 & 0.0378  & 0.0645 & 0.0702 & 0.0842 & 0.0926 \\
    100 & 0.0228 & 0.0274 & 0.0471 & 0.0510 & 0.0619 & 0.0687 \\
    200 & 0.0165 & 0.0197 & 0.0341 & 0.0370 & 0.0449 & 0.0496 \\
 \hline
\end{tabular}
 \caption{\label{fig: fig_table_Harti} Table of the dip and its associated probability to have a multimodal distribution. They have calculated the dips 9999 times on uniform data with varying sample sizes $n$. The maximum standard error is 0.001. Table reproduced from~\cite{Hartigan1985} }
\end{table}

Hartigan's dip test is a statistical method to check whether empirical statistical data is unimodal~\cite{Hartigan1985}. This test realizes a best (with the smallest deviations) fit of the empirical data with a unimodal distribution function. The largest of these deviations represents the dip, thus measuring the departure from unimodality. A large dip indicates that the empirical data distribution is more probable to be described by more than one mode. Based on the original FORTRAN algorithm of Hartigan~\cite{Hartigan1985_2}, we used a direct translation of this code into MATLAB. The latter was implemented by F. Mechler~\cite{Mechler, Mechler_2}. It was shown that the dip can be interpolated to varying sample  sizes $n$ based on the $\sqrt{n} \times (\rm dip)$. Using the values of Table.~\ref{fig: fig_table_Harti},  it is thus possible to extrapolate the value of a dip at a given sample size $n$ by using a functional form $f(n) = a/\sqrt{n} + b$, where $a$ and $b$ are free fitting parameters. In our case, for a sample size $\sim 35000$ ($\mathcal{M} \times t_{{\rm max} } $), a probability 0.99 to have more than one mode is given by a dip $f(35000) \sim 0.022$.

\bibliography{main}

\end{document}